\pdfoutput=1
\documentclass[aps,prx,reprint,superscriptaddress,longbibliography]{revtex4-2}

\usepackage{graphicx}
\usepackage{amsmath}
\usepackage{amssymb}
\usepackage[colorlinks=true,linkcolor=blue,citecolor=blue,urlcolor=blue]{hyperref}

\newcommand{\cnq}{Cr$_{1/4}$NbSe$_2$}
\newcommand{\cnt}{Cr$_{1/3}$NbSe$_2$}
\newcommand{\nbse}{NbSe$_2$}
\newcommand{\hcn}{$\mathbf{Cr}$-$\mathbf{NbSe_2}$}       
\newcommand{\cn}{Cr-NbSe$_2$}
\newcommand{\sqq}{$2\times2\mathrm{R}0^\circ$}
\newcommand{\sqt}{$\sqrt{3}\times\sqrt{3}\mathrm{R}30^\circ$}

\begin{document}

\title{Carrier-tunable RKKY magnetism in a crystalline magnet}

\author{Xiang S.W. Huang}
\affiliation{Department of Applied Physics, the University of Tokyo, Tokyo 113-8656, Japan}
\author{Bruno K. Saika}
\affiliation{Department of Applied Physics, the University of Tokyo, Tokyo 113-8656, Japan}
\affiliation{School of Physics and Astronomy, University of St Andrews, North Haugh, St Andrews, KY16 9SS, United Kingdom}
\author{Satoshi Hamao}
\affiliation{Department of Applied Physics, the University of Tokyo, Tokyo 113-8656, Japan}
\author{Yuki Majima}
\affiliation{Department of Applied Physics, the University of Tokyo, Tokyo 113-8656, Japan}
\author{Yuki Settai}
\affiliation{Department of Applied Physics, the University of Tokyo, Tokyo 113-8656, Japan}
\author{Hideki Matsuoka}
\affiliation{RIKEN Center for Emergent Matter Science (CEMS), Wako 351-0198, Japan}
\affiliation{Institute of Industrial Science, the University of Tokyo, Tokyo 153-8505, Japan}
\author{Yuki M. Itahashi}
\affiliation{RIKEN Center for Emergent Matter Science (CEMS), Wako 351-0198, Japan}
\author{Masato Sakano}
\affiliation{Department of Applied Physics, the University of Tokyo, Tokyo 113-8656, Japan}
\affiliation{Graduate School of Informatics and Engineering, The University of Electro-Communications, Chofu, Tokyo 182-8585, Japan}
\author{Taro Nakajima}
\affiliation{The Institute for Solid State Physics, University of Tokyo, Kashiwa, Chiba 277-8581, Japan}
\affiliation{RIKEN Center for Emergent Matter Science (CEMS), Wako 351-0198, Japan}
\affiliation{Institute of Materials Structure Science, High Energy Accelerator Research Organization, Tsukuba, Ibaraki 305-0801, Japan}
\author{Shinichiro Seki}
\affiliation{Department of Applied Physics, the University of Tokyo, Tokyo 113-8656, Japan}
\affiliation{Research Center for Advanced Science and Technology, University of Tokyo, Komaba, Tokyo 153-8904, Japan}
\author{Yoshihiro Iwasa}
\affiliation{Department of Applied Physics, the University of Tokyo, Tokyo 113-8656, Japan}
\affiliation{RIKEN Center for Emergent Matter Science (CEMS), Wako 351-0198, Japan}
\author{Kyoko Ishizaka}
\affiliation{Department of Applied Physics, the University of Tokyo, Tokyo 113-8656, Japan}
\affiliation{RIKEN Center for Emergent Matter Science (CEMS), Wako 351-0198, Japan}
\author{Masaki Nakano}
\email{mnakano@shibaura-it.ac.jp}
\affiliation{Department of Applied Physics, the University of Tokyo, Tokyo 113-8656, Japan}
\affiliation{RIKEN Center for Emergent Matter Science (CEMS), Wako 351-0198, Japan}
\affiliation{College of Engineering, Shibaura Institute of Technology, Tokyo 135-8548, Japan}

\begin{abstract}
In itinerant magnets governed by the Ruderman-Kittel-Kasuya-Yosida (RKKY) interaction, the exchange coupling depends on both the moment-moment distance $r$ and the Fermi wavevector $k_{\mathrm{F}}$, yet in bulk synthesis the two are tightly coupled: a change in composition typically alters both. Here, we use a thin-film approach to tune these two variables independently in a single crystalline host. Using molecular-beam epitaxy (MBE), we stabilize either \sqq{} \cnq{} or \sqt{} \cnt{} within the same \nbse{} host through separate growth windows. Controlled post-growth annealing performed across a series of temperatures then modifies the carrier density while leaving the Cr superstructure intact below a structural-transition threshold. The two as-grown phases are distinct in electronic structure, magnetic ground state, and transport. Along this annealing series, the Hall response evolves systematically while the magnetic response changes in a structurally insensitive manner, with ferromagnetic order emerging within the same \sqt{} structural class only above a critical annealing temperature, experimentally disentangling carrier density and moment geometry. The Hall magnitude and sign evolution point to a low-carrier-density system in which $k_{\mathrm{F}}$ is susceptible to modest external tuning. \cn{} thus realizes carrier-sensitive RKKY magnetism in a single crystalline host, within an MBE-plus-annealing approach extensible across the intercalated transition-metal dichalcogenide family.
\end{abstract}

\maketitle

\section{Introduction}

Itinerant electron magnetism emerges from the interplay between the lattice geometry of local moments and the conduction electrons that mediate their exchange [1,2]. The RKKY exchange coupling expresses this dependence through the oscillatory phase $2k_{F} \cdot r$ [3--6]. Independent control of $r$ and $k_{F}$ therefore offers direct access to magnetic-state selection. Dilute alloys [7], metallic multilayers [8], and bulk intercalated crystals [9--12] have each partially addressed this problem. In intercalated bulk crystals in particular, a given composition typically stabilizes a single intercalant superstructure; comparisons between phases therefore involve different samples grown under different conditions, re-entangling different variables one wishes to separate.

Intercalated transition-metal dichalcogenides (I-TMDCs) offer a structurally modular setting for this question [9--11,13--15]: the TMDC host provides an itinerant background shared by all intercalants, while ordered intercalant superstructures define a periodic moment lattice [13--15]. The van der Waals layered structure of the host TMDC in principle allows the host and intercalant sub-systems to be manipulated independently. In practice, however, traditional crystal growth methods tend to operate near thermodynamic equilibrium, in which a given composition selects a single most stable intercalant superstructure. Combined with the limited post-growth tunability of bulk crystals, this situation has prevented independent control of the two sub-systems within the same material. Achieving such control within a single crystalline host has remained an open challenge.

Using molecular-beam epitaxy (MBE), we address this issue by combining the state-of-the-art thin film growth technique with the controlled post-growth annealing in the \cn{} system. By tuning the MBE growth window, we stabilize either \cnq{} with \sqq{} intercalant ordering (Figs.~\ref{fig1}a,e) or \cnt{} with \sqt{} intercalant ordering (Figs.~\ref{fig1}b,f) in the same \nbse{} host. MBE allows precise control of the Cr/Nb flux ratio, growth temperature, Se flux, and cooling rate; the thermal history of growth, in addition to the flux ratio, plays an essential role in achieving sharp superstructure ordering. These two as-grown phases differ in their electronic, magnetic, and transport signatures. Post-growth annealing then modifies the carrier density [16,17], leaving the Cr superstructure intact below a structural-transition threshold. Along the annealing series, the Hall response evolves systematically while the magnetic response shifts in a structurally insensitive manner. The two responses disentangle moment geometry and carrier density in a single crystalline host, identifying \cn{} as a realization of carrier-sensitive RKKY magnetism.

\section{Superstructure-selective epitaxy of \texorpdfstring{\hcn{}}{Cr-NbSe2} thin films}

\begin{figure*}[t]
\includegraphics[width=\textwidth]{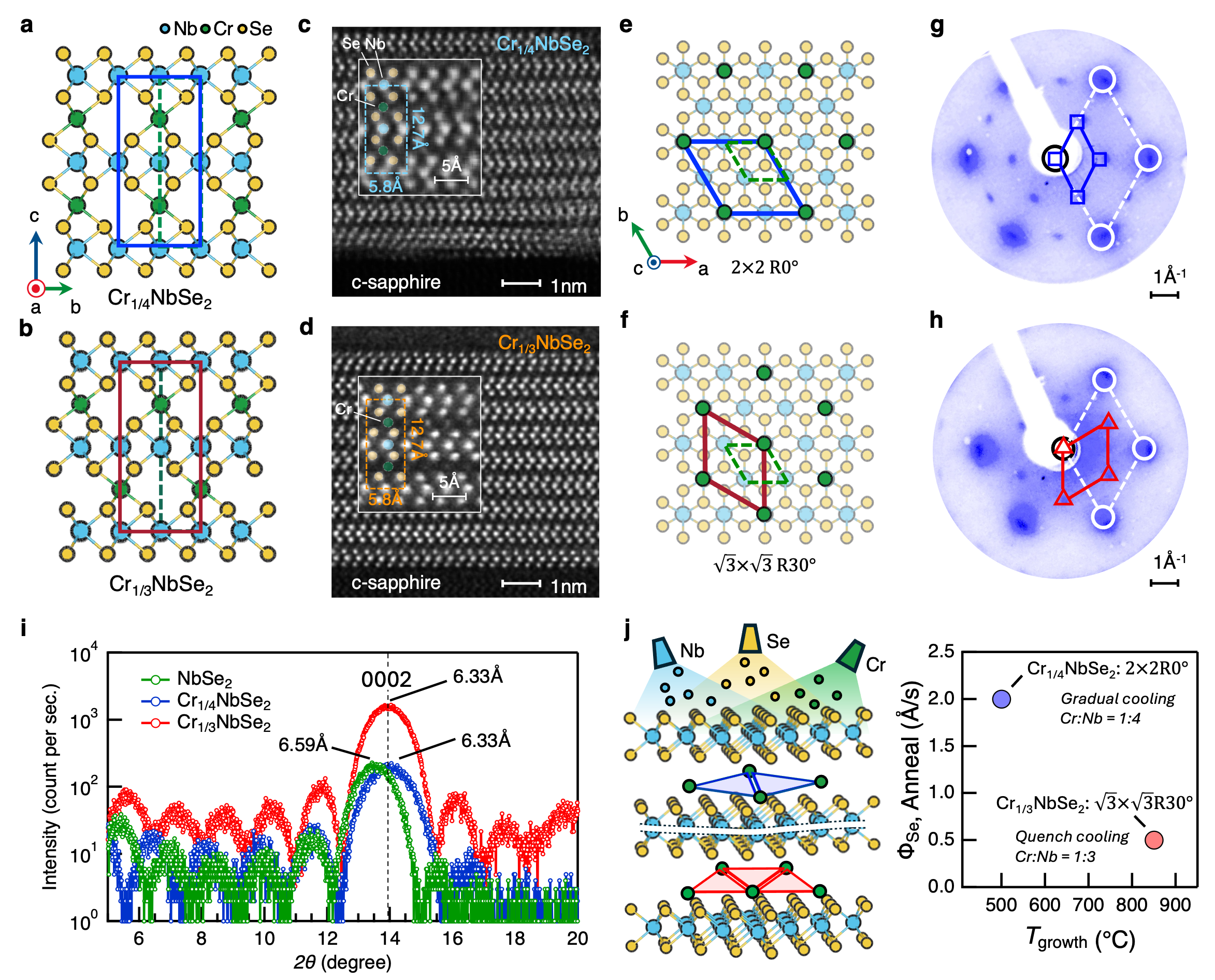}
\caption{\textbf{Superstructure-selective epitaxy of \cn{} thin films.}
\textbf{a,b,} Schematic side-view of the crystal structures of (\textbf{a}) \cnq{} and (\textbf{b}) \cnt{}. Blue, green, and yellow circles denote Nb, Cr, and Se atoms, respectively. The colored rectangles mark the supercells associated with the intercalated Cr ordering. \textbf{c,d,} Cross-sectional STEM images of (\textbf{c}) \cnq{} and (\textbf{d}) \cnt{} films grown on c-plane sapphire. The intercalated Cr atoms are visible within the van der Waals gap, and the layered epitaxial growth shows sharp interfaces (scale bars, 1~nm). The labeled distances of 5.8~\AA{} and 12.7~\AA{} correspond to the in-plane Nb-Nb diagonal and the c-axis repeat of the 2H-\nbse{} host lattice, respectively, from which the Cr-Cr separations of the two phases are obtained through the known host geometry. \textbf{e,f,} Top-view schematics of the in-plane Cr ordering: (\textbf{e}) \sqq{} in \cnq{} and (\textbf{f}) \sqt{} in \cnt{}. \textbf{g,h,} Corresponding LEED patterns showing sharp superlattice spots at the positions expected for the two Cr orderings (scale bars, 1~\AA$^{-1}$). \textbf{i,} X-ray diffraction profiles around the 0002 reflection for \nbse{}, \cnq{}, and \cnt{}. Pronounced Laue fringes indicate high out-of-plane crystallinity and uniform thickness. The 0002 peak positions give $d_{0002} = 6.59$~\AA{} for \nbse{} and $d_{0002} = 6.33$~\AA{} for both Cr-intercalated phases, consistent with the 12.7~\AA{} c-axis repeat resolved in \textbf{c,d}. \textbf{j,} A schematic illustration of superstructure-selective epitaxy of \cn{} (left) by MBE and the uncovered growth-window diagram (right). The filled circles mark the optimized MBE conditions for \cnq{} (blue, $T_{\mathrm{growth}} \approx 500$~$^\circ$C, $\Phi_{\mathrm{Se}} \approx 2$~\AA/s, gradual cooling over $\sim$30~min, Cr:Nb flux ratio 1:4) and \cnt{} (red, $T_{\mathrm{growth}} \approx 850$~$^\circ$C, $\Phi_{\mathrm{Se}} \approx 0.5$~\AA/s, rapid cooling within $\sim$5~min, Cr:Nb flux ratio 1:3). The two phases are obtained at distinct flux stoichiometries (Cr:Nb = 1:4 vs 1:3) and require distinct thermal histories (gradual vs rapid cooling) for sharp superstructure ordering.}
\label{fig1}
\end{figure*}

The \sqq{} \cnq{} phase is stabilized at a relatively low growth temperature ($T_{\mathrm{growth}} \approx 500$~$^\circ$C) under Se-rich conditions ($\Phi_{\mathrm{Se}} \approx 2$~\AA/s), followed by gradual cooling over $\sim$30~min. The \sqt{} \cnt{} phase is stabilized at a higher growth temperature ($T_{\mathrm{growth}} \approx 850$~$^\circ$C) under Se-poor conditions ($\Phi_{\mathrm{Se}} \approx 0.5$~\AA/s), followed by rapid cooling within $\sim$5~min (Fig.~\ref{fig1}j, Supplemental Material Section~A, Fig.~S1). Growth monitoring by reflection high energy electron diffraction (RHEED) shows that the Cr superstructure is not present at the end of deposition but emerges gradually during the cooling stage, confirming that the final ordering is set by the full thermal history rather than by the Cr/Nb flux ratio, which is adjusted only to match the target stoichiometries.

Cross-sectional scanning transmission electron microscope (STEM) images (Figs.~\ref{fig1}c,d) of \cnq{} and \cnt{} phases resolve the layered crystal structure with a $c$-axis lattice parameter of 12.7~\AA{} and an in-plane Nb-Nb diagonal of 5.8~\AA, giving $a_{\mathrm{NbSe_2}} \approx 3.37$~\AA{} for both phases. X-ray diffraction (XRD) profiles (Fig.~\ref{fig1}i) yield $d_{0002} = 6.33$~\AA{} for both phases, matching the STEM results; clear Laue fringes are observed under millimeter-wide x-ray illumination, attesting to wafer-scale uniformity. Low energy electron diffraction (LEED) patterns in Figs.~\ref{fig1}g,h directly resolve the \sqq{} Cr superstructure in \cnq{} and the \sqt{} superstructure in \cnt{}, in agreement with the RHEED data (Supplemental Material Section~B, Fig.~S2). The two superstructures correspond to in-plane Cr-Cr distances of $2a_{\mathrm{NbSe_2}} = 6.74$~\AA{} and $\sqrt{3}\,a_{\mathrm{NbSe_2}} = 5.84$~\AA, respectively. Together with the sharp interfaces resolved by STEM in the ultrathin regime, these observations confirm uniformly high film quality across thicknesses. Growth-window control thus enables superstructure selection between two Cr orderings within the same \nbse{} host, providing the structural basis for the experiments described below.

\section{Distinct electronic structure, magnetism, and transport signatures of the two as-grown \texorpdfstring{\hcn{}}{Cr-NbSe2} phases}

\begin{figure*}[t]
\includegraphics[width=\textwidth]{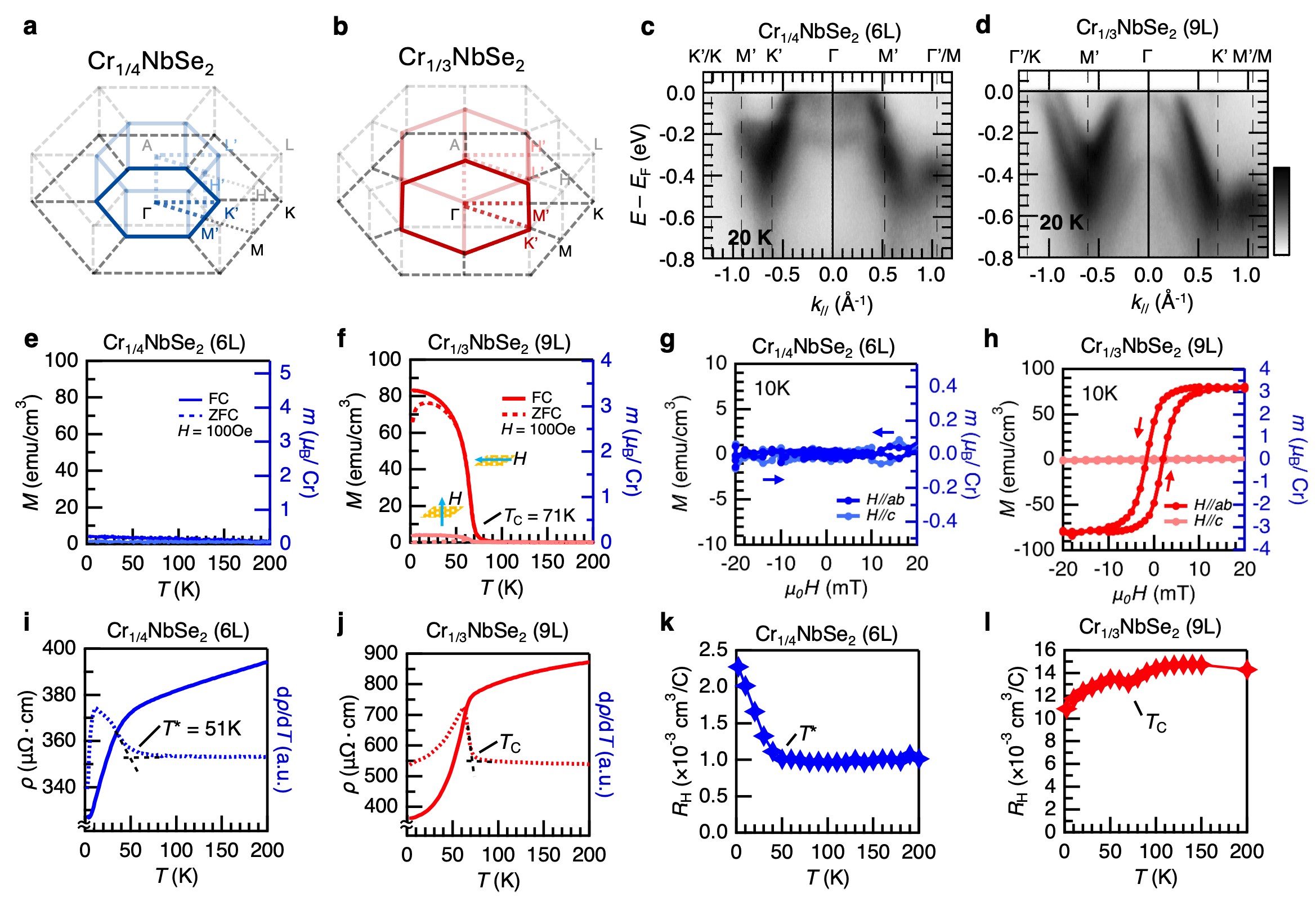}
\caption{\textbf{Distinct electronic structure, magnetism, and transport signatures of the two as-grown \cn{} phases.}
\textbf{a,b,} Reconstructed Brillouin zones of (\textbf{a}) \cnq{} and (\textbf{b}) \cnt{}, corresponding to the \sqq{} and \sqt{} Cr superstructures. The \sqq{} case folds the original high-symmetry points into $\Gamma'$, M$'$, and K$'$ without in-plane rotation, while the \sqt{} case gives a reduced zone rotated by 30$^\circ$. \textbf{c,d,} ARPES intensity plots of (\textbf{c}) 6L \cnq{} and (\textbf{d}) 9L \cnt{} films measured along representative momentum cuts at 20~K. In both phases the host \nbse{} bands shift to higher binding energy relative to pristine \nbse{}, consistent with electron donation from Cr. Additional states near $E_{F}$ appear as a relatively flat manifold in \cnq{} and as a dispersive manifold in \cnt{}. \textbf{e-h,} (\textbf{e,f}) Temperature-dependent ($M(T)$) and (\textbf{g,h}) field-dependent ($M(H)$) magnetization at 10~K of (\textbf{e,g}) \cnq{} and (\textbf{f,h}) \cnt{}. The 6L \cnq{} shows no detectable ferromagnetic signal within the present measurement resolution. The 9L \cnt{} shows a clear easy-plane ferromagnetic response with $T_{C} = 71$~K and a soft hysteresis loop. \textbf{i-l,} Temperature dependence of (\textbf{i,j}) the resistivity $\rho(T)$ and (\textbf{k,l}) Hall coefficient $R_{H}(T)$ of (\textbf{i,k}) \cnq{} and (\textbf{j,l}) \cnt{}. Both $\rho(T)$ and $R_{H}(T)$ of \cnq{} exhibit a sharp kink at a characteristic temperature $T^{*} \approx 51$~K. The resistivity of \cnt{} shows a kink near $T_{C}$, while $R_{H}(T)$ evolves more smoothly, without the sharp $T^{*}$ feature seen in \cnq{}.}
\label{fig2}
\end{figure*}

The two as-grown \cn{} phases differ in Cr content by only 1/12, yet their intercalant geometries and the resulting electronic environments are different. Figures~\ref{fig2}a,b show the reconstructed Brillouin zones of the two superstructures, which fold the original \nbse{} high-symmetry points differently: the \sqq{} case introduces no in-plane rotation, while the \sqt{} case rotates the reduced zone by 30$^\circ$. The band structures determined by the angle-resolved photoemission spectroscopy (ARPES) measurements are shown in Figs.~\ref{fig2}c,d. In both phases the host \nbse{} bands shift to higher binding energy, consistent with electron donation from Cr, but the additional states near $E_{F}$ differ in their momentum dispersions: \cnq{} hosts relatively flat bands around $\Gamma$, whereas \cnt{} shows appreciably dispersive additional bands [14,18].

\begin{figure}[t]
\includegraphics[width=\columnwidth]{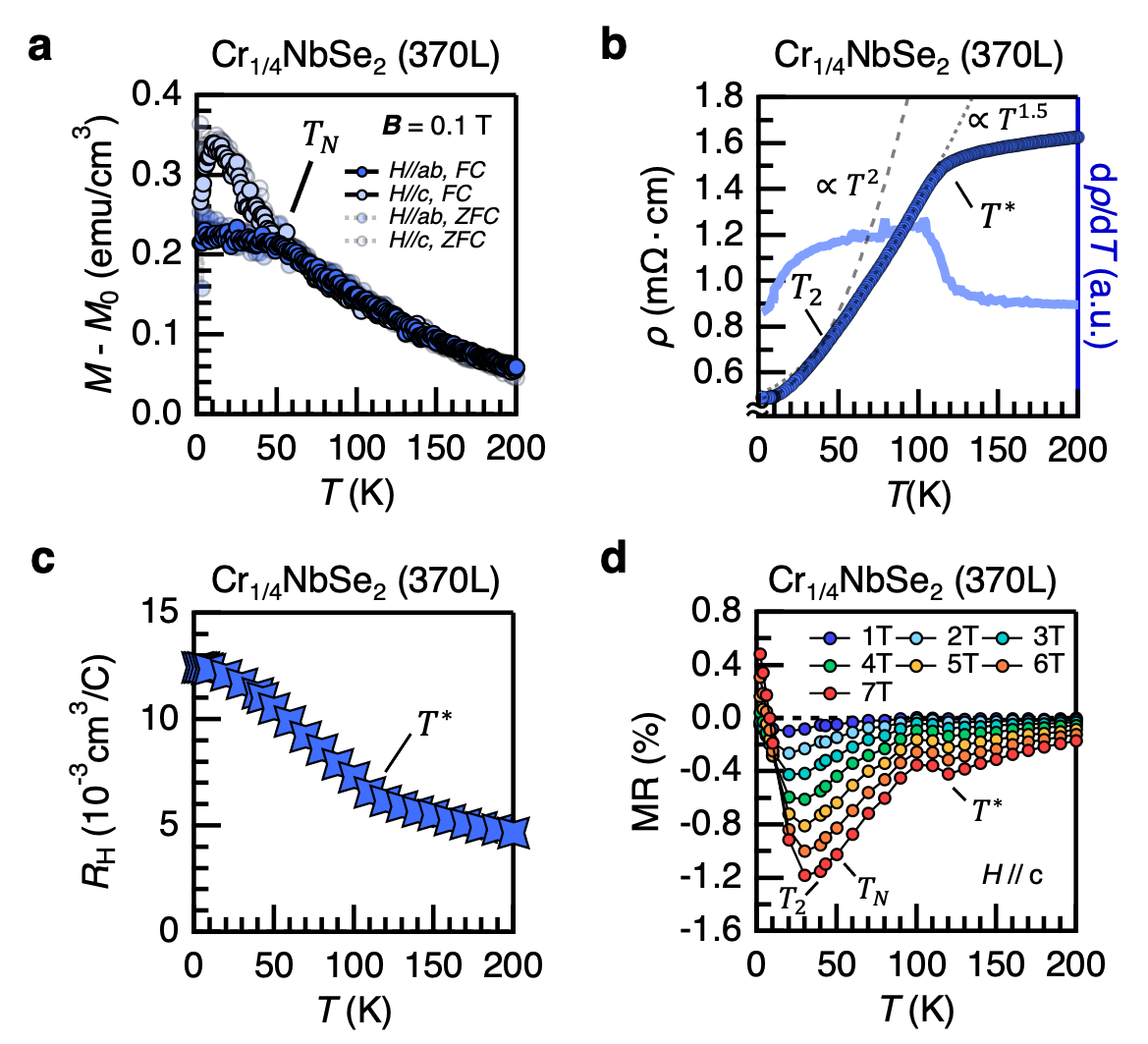}
\caption{\textbf{Magnetic and transport properties of the 370L-thick \cnq{}.}
\textbf{a,} Temperature-dependent magnetization of a 370L \cnq{} film measured with in-plane and out-of-plane fields, showing a clear N\'eel-type peak at $T_{N}$. \textbf{b,} Temperature dependence of the resistivity $\rho(T)$ together with its temperature derivative $d\rho/dT$. Three phenomenological scales are indicated: a low-temperature $\rho \propto T^{2}$ regime up to $T_{2}$, an intermediate sub-quadratic regime approximated by $\rho \propto T^{1.5}$, and the higher characteristic temperature $T^{*}$. \textbf{c,} Temperature dependence of the Hall coefficient $R_{H}(T)$, which evolves strongly across the sub-quadratic regime. \textbf{d,} Magnetoresistance $MR(T,H)$ for $H$//c at fields from 1 to 7~T, resolving two distinct features: a field-dependent negative response with minimum near $T_{2} \lesssim T_{N}$ and a higher-temperature border at $T^{*}$, and providing single-panel visualization of the three transport-magnetic scales.}
\label{fig3}
\end{figure}

The magnetic ground states are also different. For 9-layer \cnt{}, $M(T)$ in Fig.~\ref{fig2}f exhibits a ferromagnetic transition at Curie temperature $T_{C} = 71$~K with strong in-plane anisotropy, and the 10~K $M(H)$ loop in Fig.~\ref{fig2}h shows a soft hysteresis for in-plane fields, identifying it as an easy-plane ferromagnet. For 6-layer \cnq{}, both $M(T)$ and $M(H)$ (Figs.~\ref{fig2}e,g) are featureless within the present measurement resolution due to the limited magnetic volume; to determine the ground state, we grew a 370-layer \cnq{} film as a reference sample. In the 370L film, $M(T)$ in Fig.~\ref{fig3}a exhibits a N\'eel-type peak at $T_{N} \approx 50$~K, with the $H$//ab response substantially larger than $H$//c, identifying it as an easy-plane antiferromagnet (AFM). A bulk \cnq{} sample shows similar behavior with $T_{N} \approx 65$~K (Supplemental Material Section~C, Fig.~S3a), suggesting that our \cnq{} films share the same magnetic ground state. The neutron scattering experiments on the same bulk sample reveal a predominantly 120$^\circ$ in-plane AFM character as the magnetic ground state of \cnq{} (Supplemental Fig.~S3e), which is consistent with the early reports of AFM order in \cnq{} [19,20].

Transport responses are different as well. Both $\rho(T)$ and $R_{H}(T)$ of 6-layer \cnq{} exhibit a sharp kink at a characteristic temperature $T^{*} \approx 51$~K (Figs.~\ref{fig2}i,k), without a detectable anomalous Hall component (Supplemental Fig.~S9a). For 9-layer \cnt{}, $\rho(T)$ shows a kink at $T_{C}$ (Fig.~\ref{fig2}j), $R_{H}(T)$ evolves smoothly (Fig.~\ref{fig2}l), and an anomalous Hall component emerges below $T_{C}$ (Supplemental Fig.~S9b).

The 370L film reveals additional features in the \cnq{} transport response (Figs.~\ref{fig3}b-d). The resistivity follows a power-law form
\begin{equation}
\rho(T) - \rho_{0} \propto T^{\alpha},
\label{eq1}
\end{equation}
with two distinct exponents along the temperature axis: a canonical Fermi-liquid value $\alpha = 2$ below $T_{2}$, and a sub-quadratic value $\alpha \approx 1.5$ between $T_{2}$ and $T^{*}$, with $T_{2}$ lying slightly below $T_{N}$. The Hall coefficient $R_{H}(T)$ evolves strongly across this sub-quadratic regime.

Magnetoresistance (MR) measurements with $H$//c (Fig.~\ref{fig3}d) display two distinct features: a field-dependent negative response with a minimum near $T_{2} \lesssim T_{N}$, and a higher-temperature border at $T^{*}$. This single-measurement visualization independently confirms the AFM transition and the sub-quadratic transport onset identified in $M(T)$ and $\rho(T)$. Complementary magnetoresistance measurements on the 6L film in three field-current geometries confirm in-plane isotropy of the magnetic response, an anisotropy onset at $T^{*} \approx 50$~K, and a sign change in MR($H$//ab) within the antiferromagnetic phase, providing transport-based identification of the magnetic ordering scale in the ultrathin limit where $M(T)$ is featureless (Supplemental Material Section~D, Fig.~S5).

\cnq{} thus shows three temperature scales: $T_{2} \lesssim T_{N} < T^{*}$, combining an antiferromagnetic ground state with a higher-temperature sub-quadratic transport regime.

\cnq{} and \cnt{} are therefore qualitatively distinct in the electronic, magnetic, and transport signatures despite their near-identical Cr content. We next consider whether this structure-magnetism correspondence is one-to-one when the carrier density is varied.

\section{Controlled post-growth annealing of \texorpdfstring{\hcn{}}{Cr-NbSe2}}

\begin{figure*}[t]
\includegraphics[width=\textwidth]{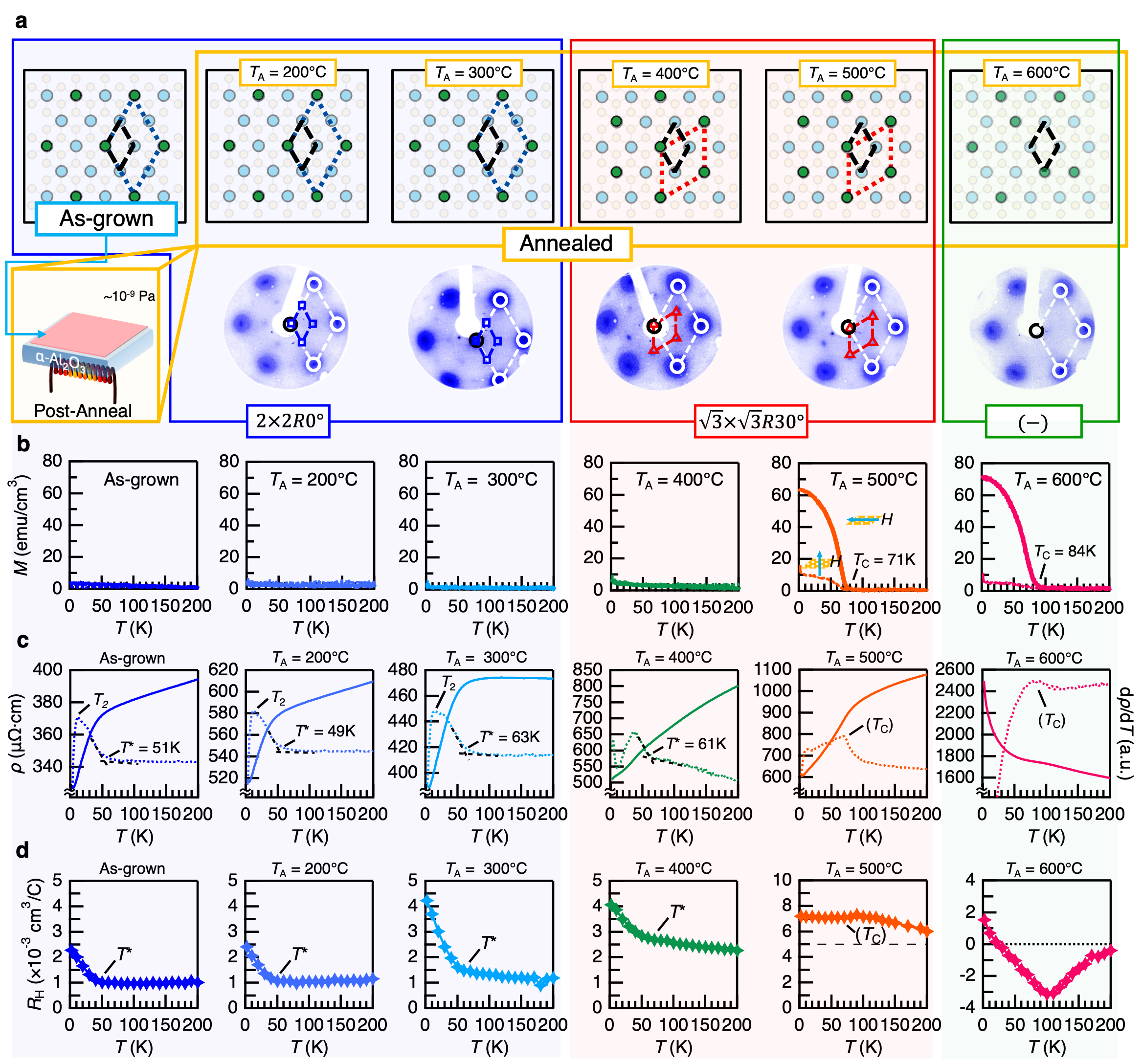}
\caption{\textbf{Controlled post-growth annealing of \cnq{}: non-bijective evolution of structure, magnetism, and transport.}
\textbf{a,} Evolution of the in-plane Cr superstructure in \cnq{} upon post-growth annealing. The as-grown 6L \cnq{} film was cut into pieces and annealed at $T_{A} = 200$~$^\circ$C, 300~$^\circ$C, 400~$^\circ$C, 500~$^\circ$C, and 600~$^\circ$C for 1~h. The LEED-resolved superstructure evolves from \sqq{} (retained up to $T_{A} = 300$~$^\circ$C) to \sqt{} ($T_{A} = 400$~$^\circ$C and 500~$^\circ$C), and finally to an unresolved state at $T_{A} = 600$~$^\circ$C. \textbf{b,} $M(T)$ curves taken under field-cooling at 100~Oe with in-plane and out-of-plane fields. In-plane and out-of-plane data almost overlap for the as-grown, 200~$^\circ$C, 300~$^\circ$C, and 400~$^\circ$C samples, and no ferromagnetic signal is detected for those samples. Easy-plane ferromagnetism emerges at $T_{A} = 500$~$^\circ$C with $T_{C} = 71$~K, and persists at $T_{A} = 600$~$^\circ$C with $T_{C} = 84$~K. The 400~$^\circ$C and 500~$^\circ$C samples share the \sqt{} structural assignment but differ qualitatively in magnetic ground state, and the 600~$^\circ$C sample remains ferromagnetic in the absence of long-range Cr order. \textbf{c,} $\rho(T)$ curves together with $d\rho/dT$ for the same annealing series. The as-grown, 200~$^\circ$C, 300~$^\circ$C, and 400~$^\circ$C samples retain a kink-like feature at $T^{*}$, the 500~$^\circ$C sample develops a ferromagnetic-type kink near $T_{C}$, and the 600~$^\circ$C sample shows an insulating-like temperature dependence. \textbf{d,} $R_{H}(T)$ for the same series, which evolves systematically with $T_{A}$, in contrast to the step-like magnetic response in \textbf{b}. The sign change of $R_{H}$ at $T_{A} = 600$~$^\circ$C marks a qualitative change in the carrier response.}
\label{fig4}
\end{figure*}

We apply post-growth vacuum annealing to the as-grown 6L \cnq{} film as a secondary tuning procedure. Pieces cut from the same wafer were annealed for 1~h at $T_{A} = 200$, 300, 400, 500, and 600~$^\circ$C, with LEED taken before and after to track the structural evolution (Supplemental Figs.~S6 and S7). After Se de-capping, all samples retain the \sqq{} superstructure (Supplemental Fig.~S7). Upon 1~h of annealing, this \sqq{} ordering persists at $T_{A} = 200$ and 300~$^\circ$C, transforms to \sqt{} at 400 and 500~$^\circ$C, and becomes diffractively unresolved at 600~$^\circ$C (Fig.~\ref{fig4}a and Supplemental Fig.~S7). Raman spectra across the same series (Supplemental Fig.~S8) evolve consistently with the LEED-resolved change, indicating that the structural reorganization affects the full film thickness, instead of being a surface effect only. The Raman results are also consistent with earlier observation of intercalant superlattice formation in bulk Fe-intercalated \nbse{} [21]. The present \cn{} spectra display analogous superlattice-derived phonon signatures across the $2\times2$ and $\sqrt{3}\times\sqrt{3}$ ordering windows, consistent with this earlier methodology.

The magnetic response does not follow this structural evolution. Figure~\ref{fig4}b shows $M(T)$ curves under 100~Oe both in-plane and out-of-plane field cooling. No ferromagnetic signal is detected for the as-grown, 200, 300, or 400~$^\circ$C samples. Ferromagnetism emerges abruptly at $T_{A} = 500$~$^\circ$C ($T_{C} = 71$~K) and persists at 600~$^\circ$C ($T_{C} = 84$~K). The 400 and 500~$^\circ$C samples change to the \sqt{} structure, as independently resolved by LEED and Raman, yet exhibit different magnetic ground states: non-ferromagnetic at 400~$^\circ$C vs. ferromagnetic with $T_{C} = 71$~K at 500~$^\circ$C. The 600~$^\circ$C sample remains ferromagnetic in the absence of any LEED-resolvable Cr ordering. The 500~$^\circ$C sample's $T_{C} = 71$~K matches that of as-grown 9L \cnt{}, and its saturation magnetization is approximately 3/4 that of the 9L \cnt{} reference.

The 3/4 ratio identifies the Cr sublattice as the source of the spontaneous magnetization and is quantitatively consistent with conservation of the precursor Cr content distributed into a 3/4 \cnt{} + 1/4 \nbse{} phase mixture. The Cr superstructure therefore does not uniquely determine the magnetic ground state in this annealing series: a non-bijective relation between Cr superstructure and magnetic ground state in \cn{}.

Transport results provide the main observables that distinguish samples across the series (Figs.~\ref{fig4}c,d). The as-grown, 200, 300, and 400~$^\circ$C samples retain a sharp kink at $T^{*}$ in both $\rho(T)$ and $R_{H}(T)$. The 500~$^\circ$C sample, corresponding to the $M(T)$, develops a kink near $T_{C}$, and $R_{H}$ increases substantially relative to the lower-$T_{A}$ samples. The 600~$^\circ$C sample shows insulating-like $\rho(T)$ and an $R_{H}(T)$ where $R_{H}$ changes sign with temperature. We also confirm the anomalous Hall effect for the 500 and 600~$^\circ$C samples (Supplemental Figs.~S9c,d). $R_{H}$ increases through the as-grown to 500~$^\circ$C samples and turns negative at high temperatures at 600~$^\circ$C, indicating progressive doping of the itinerant background (Supplemental Fig.~S10).

The temperature-dependent annealing series of Fig.~\ref{fig4} already shows that carrier and structural responses can decouple along the way: between as-grown and $T_{A} = 300$~$^\circ$C, the \sqq{} superstructure is preserved, while $\rho(T)$ and $R_{H}(T)$ evolve systematically (Supplemental Material Section~E). This carrier evolution may be caused by Se-vacancy formation [16,17]. Under the ultrahigh-vacuum conditions of the annealing chamber, Se evaporates preferentially from the surface at elevated $T_{A}$, donating electrons to the \nbse{} host. This scenario fits the monotonic increase of $R_{H}$ at low $T_{A}$ and its sign change at 600~$^\circ$C.

\section{Phase diagram and carrier-sensitive RKKY magnetism in \texorpdfstring{\hcn{}}{Cr-NbSe2}}

\begin{figure*}[t]
\includegraphics[width=\textwidth]{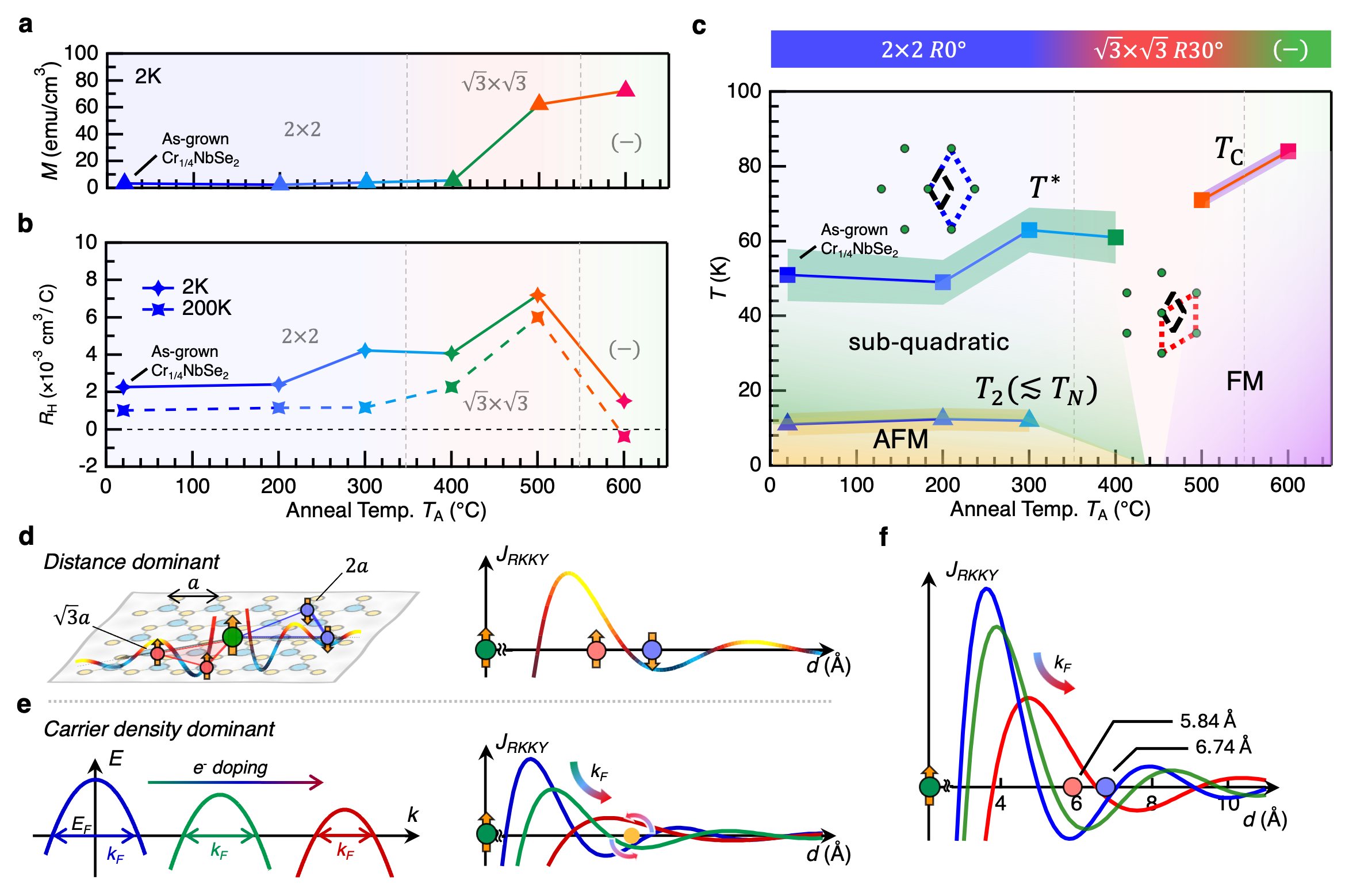}
\caption{\textbf{Phase diagram and carrier-sensitive RKKY magnetism in \cn{}.}
\textbf{a,b,} Annealing-temperature dependence of the magnetization $M$ at 2~K (\textbf{a}), and that of the Hall coefficient $R_{H}$ at 2~K and 200~K (\textbf{b}) across the \cnq{} annealing series. The shaded regions indicate the Cr superstructures assigned by LEED: \sqq{} (blue), \sqt{} (red), and the unresolved state (green). Magnetism jumps between discrete states whose boundaries are not aligned with the structural-class boundaries, while the Hall response evolves systematically across the same annealing trajectory. \textbf{c,} Phase diagram in the ($T_{A}$, $T$) plane. $T^{*}$ marks the upper boundary of the sub-quadratic transport regime, $T_{2}$ marks the crossover into the low-temperature $\rho \propto T^{2}$ regime inside the antiferromagnetic ground state, and the $T_{C}$ points mark the ferromagnetic transitions of the 500~$^\circ$C and 600~$^\circ$C samples. The color-coded band at the top indicates the Cr-superstructure class along the annealing axis. \textbf{d,e,} Limiting RKKY pictures in which the exchange coupling is tuned predominantly by the local-moment spacing $r$ (\textbf{d}) or by carrier-induced changes of $k_{F}$ (\textbf{e}). \textbf{f,} Schematic of the combined $J_{\mathrm{RKKY}}(r, k_{F})$ for \cn{}. The Cr superstructure fixes the real-space distance axis at the discrete Cr-Cr values 6.74~\AA{} for \cnq{} (blue marker) and 5.84~\AA{} for \cnt{} (red marker), while annealing tunes $k_{F}$ along the electronic axis (arrow). Panels \textbf{d-f} are schematic illustrations consistent with the observed trends; quantitative calibration to the \cn{} band structure is not attempted in this work.}
\label{fig5}
\end{figure*}

Figures~\ref{fig5}a and \ref{fig5}b summarize the annealing-temperature dependence of the magnetization $M$ at 2~K and the Hall coefficient $R_{H}$ at 2~K and 200~K, respectively. The shaded regions indicate the Cr superstructure assigned by LEED. Magnetization remains at the detection floor across the as-grown, 200, 300, and 400~$^\circ$C samples, jumps abruptly at $T_{A} = 500$~$^\circ$C, and stays elevated at 600~$^\circ$C. The Hall coefficient evolves systematically through the same series, with no comparable step. The contrast in Figs.~\ref{fig5}a,b makes carrier density visible as a tunable axis along the same annealing trajectory on which the magnetic ground state changes discretely.

Figure~\ref{fig5}c organizes these observations as a phase diagram in the ($T_{A}$, $T$) plane. The three transport-defined temperature scales of \cnq{} introduced above, $T^{*}$, $T_{2}$, and $T_{N}$ (resolved in the 370L reference film), are tracked across the annealing series, together with the $T_{C}$ points of the 500 and 600~$^\circ$C samples. The structural class along the $T_{A}$ axis is indicated by the color band at the top. The diagram resolves three regions: an antiferromagnetic ground state at low $T_{A}$ bounded above by $T_{N}$ and capped at higher temperature by the $T^{*}$-bounded sub-quadratic regime; a ferromagnetic region at high $T_{A}$ anchored by $T_{C}$; and an intermediate window at $T_{A} = 400$~$^\circ$C in which the superstructure has changed but ferromagnetism has not yet emerged.

The trajectory through Fig.~\ref{fig5}c invites a unified interpretation in terms of RKKY exchange coupling. The asymptotic form of the RKKY coupling between two local moments separated by $r$ in an itinerant background of Fermi wavevector $k_{F}$ is
\begin{equation}
J_{\mathrm{RKKY}}(r, k_{F}) \propto \frac{F(2k_{F}r)}{r^{n}},
\label{eq2}
\end{equation}
where $F$ is an oscillatory function and the exponent $n$ depends on the dimensionality and band structure of the mediating electronic system [3--5,22]. The argument $2k_{F}r$ enters through both $r$ and $k_{F}$, so a sufficient change in either variable can carry the exchange through a sign reversal.

The limiting pictures provide direct intuition for each variable. In the distance-dominant limit (Fig.~\ref{fig5}d), $k_{F}$ is held fixed and $r$ is varied; this is the picture realized in metallic multilayers, where moments and conduction carriers reside in spatially separated sub-systems and the exchange is tuned by spacer thickness [8]. In the carrier-density-dominant limit (Fig.~\ref{fig5}e), $r$ is held fixed and $k_{F}$ is varied, so that $J_{\mathrm{RKKY}}$ at a fixed spacing changes sign as $k_{F}$ evolves. \cn{} as probed by our annealing series sits in neither pure limit (Fig.~\ref{fig5}f): the Cr superstructure fixes $r$ at one of two discrete values ($2a_{\mathrm{NbSe_2}} = 6.74$~\AA{} for \cnq{} and $\sqrt{3}\,a_{\mathrm{NbSe_2}} = 5.84$~\AA{} for \cnt{}), while annealing tunes $k_{F}$ along the electronic axis. Magnetic-state selection follows from the joint position on the ($r$, $k_{F}$) plane.

In the annealing series, the Hall coefficients are large ($\sim$$10^{-3}$~cm$^{3}$/C), strongly annealing-temperature dependent with sign reversal in one sample, and non-linear Hall responses are observed for the as-grown samples (Supplemental Material Section~D, Fig.~S4). These transport features indicate a low-carrier-density, semimetal-like multicarrier system, in which modest perturbation can shift $k_{F}$ and affect the RKKY exchange oscillation significantly.

\section{Discussion}

The interplay between local magnetic moments and itinerant conduction electrons has long served as one of the central questions in condensed matter physics, organizing fields ranging from dilute magnetic alloys [7] and metallic multilayers [8] to intercalated crystalline hosts [9--13]. \cn{} as developed here adds a distinct setting to this line: a single crystalline host in which the magnetic moments occupy a structurally resolved periodic lattice, the itinerant background is a uniform conduction medium shared by both phases, and the carrier density is independently tunable by post-growth annealing. Unlike disordered alloys where the positions of the magnetic moments are random, multilayer geometries where moments and carriers reside in spatially separated sub-systems, or bulk intercalates whose structure and carrier density move together with composition, \cn{} epitaxial films provide an ideal platform that keeps the moment lattice and the carrier axis as separately controllable experimental variables within the same material. Carrier-induced magnetic transitions have also been reported in stoichiometric low-carrier-density Eu-based pnictides such as EuCd$_2$As$_2$ [23] and EuZn$_2$P$_2$ [24], but those systems sit below the validity threshold of the RKKY framework and operate in a regime where super-exchange competes with carrier-mediated coupling [24].

The central observation of this work is the non-bijective correspondence between Cr superstructure and magnetic ground state demonstrated in Fig.~\ref{fig4}: two samples that share the \sqt{} superstructure (the 400~$^\circ$C and 500~$^\circ$C annealed samples) exhibit qualitatively different magnetic ground states. This rules out a one-to-one structure-magnetism mapping in this annealing series and identifies the carrier density, which evolves systematically across the same trajectory, as the operative axis selecting the magnetic ground state at fixed moment geometry.

One related observation has been reported very recently in bulk V$_{1/3}$NbS$_2$, where two superlattices differing in out-of-plane intercalant periodicity coexist within the same single crystal and give rise to distinct magnetic ground states [25]. That work established that an I-TMDC host can support multiple ordered magnetic phases, but the bulk synthesis yields these phases as a mixed sample, and the carrier density is hard to tune independently of the structural composition. Our work brings this observation into the thin-film regime and resolves both limitations: the two \cn{} superstructures are obtained as single-phase epitaxial films through MBE growth-window selection, and the post-growth annealing protocol provides a separate carrier-density tuning axis. The thin-film implementation therefore extends the in-host structural multiplicity already noted in bulk to a setting where the lattice and carrier axes can be addressed independently.

\section{Conclusion}

We have established \cn{} as a crystalline platform in which moment geometry and carrier density are independently tunable, through the combination of growth-window-selective MBE and controlled post-growth annealing. Along the annealing series, the Hall response evolves systematically while the magnetic response shifts in a structurally insensitive manner, experimentally disentangling the two variables in a single crystalline host. The Hall response suggests a low-carrier-density itinerant background, placing \cn{} in an RKKY regime where the magnetic ground state is shaped by the joint ($r$, $k_{F}$) coordinate; the MBE-plus-annealing approach demonstrated here provides a generalizable route to explore this regime across the I-TMDC family.

\section{Methods}

\subsection*{Substrate preparation and characterization}
All the thin film samples were grown on sapphire ($\alpha$-Al$_2$O$_3$) (001) substrates (SHINKOSHA). The sapphire substrates were annealed at 1000~$^\circ$C in air for 3 hours before transferring to the MBE chamber. The substrate pre-anneal process is for forming an atomically-flat surface with a step and terrace structure (Fig.~S11). The terrace width is about 1~$\mu$m with the step height of around 3~\AA, characterized by an atomic force microscopy (Hitachi High-Tech, AFM5100N).

\subsection*{Thin film growth and structural characterizations}
Thin film growth was proceeded in the MBE chamber (EIKO Engineering) with a base pressure $\sim 1 \times 10^{-7}$~Pa. Se and Cr were supplied by standard Knudsen cells, while Nb was supplied by an electron beam evaporator. By altering parameters with different growth recipes, we were able to selectively grow the \cnq{} and \cnt{} epitaxial thin films. Details of the growth process are explained and discussed in Supplemental Material Sections~A and B. We used a reflection high energy electron diffraction (RHEED) system (EIKO Engineering) for the \textit{in-situ} growth monitoring. After the growth, $\sim$1~$\mu$m-thick Se capping layers were formed to protect the thin films from oxidization. The crystallinities of the obtained films were characterized by a four-circle x-ray diffractometer (PANalytical, Empyrean).

\subsection*{Post-growth annealing and characterizations by LEED and ARPES}
Post-growth annealing was proceeded in an ultra-high vacuum system with a base pressure $\sim 1 \times 10^{-8}$~Pa. The in-plane superstructures of Cr were characterized by a low energy electron diffraction (LEED) system (OCI Vacuum Microengineering) equipped with the same vacuum chamber. The angle-resolved photoemission spectroscopy (ARPES) measurements were performed in the same chamber when needed. After the annealing and LEED/ARPES characterizations, the samples were capped by thick-enough Se and took out from the chamber for further measurements. ARPES measurement with 21.2~eV photon energy was performed using a VUV5000 He-discharge lamp and DA30 hemispherical electron analyzer (Scienta Omicron). The total energy resolution was set to 20~meV.

\subsection*{Magnetization and transport measurements}
Magnetization measurements were performed by Magnetic Property Measurement System 3 (Quantum Design, MPMS3). The electrical transport properties were measured by Physical Property Measurement System (Quantum Design, PPMS). All the samples were cut into Hall-bar shape before transport measurements by mechanical scratching to define the channel regions, which were typically a few hundred micrometers.

\subsection*{Raman spectroscopy}
Room-temperature Raman spectra of the annealed \cnq{} films were collected using a 532~nm excitation laser by LabRAM Odyssey spectrometer (HORIBA). The samples were measured through the Se capping layer.

\subsection*{Neutron scattering measurements}
Neutron scattering measurements on a bulk single crystal of \cnq{} were carried out at the POlarized Neutron Triple-Axis spectrometer PONTA installed at the 5G beamhole of Japan Research Reactor 3 (JRR-3) [26]. The incident neutrons with the energy of 34.05~meV was obtained by a pyrolytic graphite (PG) monochromator. The spectrometer was operated in a two-axis diffraction mode with the horizontal beam collimation of open-80$'$-80$'$. A single crystal of \cnq{} with a mass of 10~mg was mounted in an Al sample cell with a small amount of $^4$He gas for thermal exchange. The horizontal scattering plane was selected to be the ($H$, $H$, $L$) plane. The sample cell was attached to the cold head of a $^4$He closed-cycle refrigerator and was cooled to low temperatures.

\section*{Data availability}
The data that support the findings of this study are available from the corresponding author upon reasonable request.

\begin{acknowledgments}
We are grateful to S. Jiao, L. Zhang, M. Naritsuka, T. Ideue for valuable discussions. This work was supported by Grants-in-Aid for Scientific Research (Grant No.~22H01949, 25H00846, 21K13888, 24H01212, 21H05235, 24K01285) from the Japan Society for the Promotion of Science (JSPS), and by PRESTO (Grant No.~JPMJPR20AC) and by FOREST (Grant No.~JPMJFR223W) from Japan Science and Technology Agency. X.S.W.H. was supported by the World-Leading Innovative Graduate Study Program for the Quantum and Semiconductor Science and Technology Fellowship Program (WINGS-QSTEP). H.M. was partly supported by research granted from Murata Science and Education Foundation. M.N. was partly supported by The Mitsubishi Foundation and by SIT Supporting Program for Innovative Research (S-SPIRE). The neutron scattering experiments at PONTA were carried out along the proposals (No.~23401) and partly supported by the institute for solid state physics of the University of Tokyo. This study was carried out by the joint research of the Cryogenic Research Center, the University of Tokyo.
\end{acknowledgments}

\section*{Author contributions}
X.S.W.H. and Y.M. grew and characterized the samples, performed magnetization measurements and analyzed the data. B.K.S., S.H., and M.S. performed the post-growth annealing experiments, LEED measurements, and ARPES measurements. Y.S., S.S. grew the bulk sample, T.N. performed the neutron diffraction measurement. X.S.W.H., Y.M., H.M., and Y.M.I. performed magnetization and transport measurements, and analyzed the data. M.N., K.I., and Y.I. supervised the study. X.S.W.H. and M.N. wrote the manuscript. All the authors discussed the results and commented on the manuscript.

\section*{Competing interests}
The authors declare no competing interests.

\section*{Correspondence}
Correspondence and requests for materials should be addressed to M.N.


\clearpage


\begin{thebibliography}{99}

\bibitem{Anderson1978} P. W. Anderson, Local moments and localized states, Rev. Mod. Phys. \textbf{50}, 191-201 (1978).
\bibitem{Lonzarich2017} G. G. Lonzarich, D. Pines, and Y.-F. Yang, Toward a new microscopic framework for Kondo lattice materials, Rep. Prog. Phys. \textbf{80}, 024501 (2017).
\bibitem{Ruderman1954} M. A. Ruderman and C. Kittel, Indirect Exchange Coupling of Nuclear Magnetic Moments by Conduction Electrons, Phys. Rev. \textbf{96}, 99-102 (1954).
\bibitem{Kasuya1956} T. Kasuya, A Theory of Metallic Ferro- and Antiferromagnetism on Zener's Model, Prog. Theor. Phys. \textbf{16}, 45-57 (1956).
\bibitem{Yosida1957} K. Yosida, Magnetic Properties of Cu-Mn Alloys, Phys. Rev. \textbf{106}, 893-898 (1957).
\bibitem{Coleman2015} P. Coleman, \textit{Introduction to Many-Body Physics} (Cambridge University Press, 2015).
\bibitem{Mydosh2015} J. A. Mydosh, Spin glasses: Redux: An updated experimental/materials survey, Rep. Prog. Phys. \textbf{78}, 052501 (2015).
\bibitem{Parkin1990} S. S. P. Parkin, N. More, and K. P. Roche, Oscillations in exchange coupling and magnetoresistance in metallic superlattice structures: Co/Ru, Co/Cr, and Fe/Cr, Phys. Rev. Lett. \textbf{64}, 2304-2307 (1990).
\bibitem{Wang2012} Q. H. Wang, K. Kalantar-Zadeh, A. Kis, J. N. Coleman, and M. S. Strano, Electronics and optoelectronics of two-dimensional transition metal dichalcogenides, Nat. Nanotechnol. \textbf{7}, 699-712 (2012).
\bibitem{Wang2020} Z. Wang, R. Li, C. Su, and K. P. Loh, Intercalated phases of transition metal dichalcogenides, SmartMat \textbf{1}, e1013 (2020).
\bibitem{Xie2022} L. S. Xie, S. Husremovi\'c, O. Gonzalez, I. M. Craig, and D. K. Bediako, Structure and magnetism of iron- and chromium-intercalated niobium and tantalum disulfides, J. Am. Chem. Soc. \textbf{144}, 9525-9542 (2022).
\bibitem{Ko2011} K.-T. Ko \textit{et al.}, RKKY Ferromagnetism with Ising-Like Spin States in Intercalated Fe$_{1/4}$TaS$_2$, Phys. Rev. Lett. \textbf{107}, 247201 (2011).
\bibitem{Parkin1980} S. S. P. Parkin and R. H. Friend, 3d transition-metal intercalates of the niobium and tantalum dichalcogenides. I. Magnetic properties, Philos. Mag. B \textbf{41}, 65-93 (1980).
\bibitem{Saika2022} B. K. Saika \textit{et al.}, Signature of topological band crossing in ferromagnetic Cr$_{1/3}$NbSe$_2$ epitaxial thin film, Phys. Rev. Research \textbf{4}, L042021 (2022).
\bibitem{Togawa2012} Y. Togawa \textit{et al.}, Chiral Magnetic Soliton Lattice on a Chiral Helimagnet, Phys. Rev. Lett. \textbf{108}, 107202 (2012).
\bibitem{Bonilla2020} M. Bonilla \textit{et al.}, Compositional phase change of early transition metal diselenide VSe$_2$ and TiSe$_2$ ultra-thin films by postgrowth annealing, Adv. Mater. Interfaces \textbf{7}, 2000497 (2020).
\bibitem{Andreeva2012} O. N. Andreeva, I. S. Braude, and A. A. Mamalui, Selenium vacancies and their effect on the fine structure of NbSe$_2$ quasi-two-dimensional single crystals, Phys. Met. Metallogr. \textbf{113}, 888-892 (2012).
\bibitem{Xie2023} L. S. Xie \textit{et al.}, Comparative electronic structures of the chiral helimagnets Cr$_{1/3}$NbS$_2$ and Cr$_{1/3}$TaS$_2$, Chem. Mater. \textbf{35}, 7239-7251 (2023).
\bibitem{Hulliger1970} F. Hulliger and E. Pobitschka, On the magnetic behavior of new 2H-NbS$_2$-type derivatives, J. Solid State Chem. \textbf{1}, 117-119 (1970).
\bibitem{Yamaoka2026} R. Yamaoka \textit{et al.}, Magnetic order and excitations in the magnetically intercalated van der Waals material Cr$_{1/4}$NbSe$_2$, arXiv:2604.07117 (2026).
\bibitem{Erodici2023} M. P. Erodici \textit{et al.}, Bridging Structure, Magnetism, and Disorder in Iron-Intercalated Niobium Diselenide, Fe$_x$NbSe$_2$, below $x$ = 0.25, J. Phys. Chem. C \textbf{127}, 9787-9795 (2023).
\bibitem{Aristov1997} D. N. Aristov, Indirect RKKY interaction in any dimensionality, Phys. Rev. B \textbf{55}, 8064-8066 (1997).
\bibitem{Jo2020} N. H. Jo \textit{et al.}, Manipulating of magnetism in the topological semimetal EuCd$_2$As$_2$, Phys. Rev. B \textbf{101}, 140402(R) (2020).
\bibitem{Chen2024} X. Y. Chen \textit{et al.}, Carrier-induced transition from antiferromagnetic insulator to ferromagnetic metal in the layered phosphide EuZn$_2$P$_2$, Phys. Rev. B \textbf{109}, L180410 (2024).
\bibitem{Fender2025} S. S. Fender \textit{et al.}, Unconventional superlattice ordering in intercalated transition metal dichalcogenide V$_{1/3}$NbS$_2$, J. Am. Chem. Soc. \textbf{147}, 32315-32320 (2025).
\bibitem{Nakajima2024} T. Nakajima \textit{et al.}, Polarized and Unpolarized Neutron Scattering for Magnetic Materials at the Triple-axis Spectrometer PONTA in JRR-3, J. Phys. Soc. Jpn. \textbf{93}, 091002 (2024).

\end{thebibliography}
\end{document}


\title{Supplemental Material for ``Carrier-tunable RKKY magnetism in a crystalline magnet''}

\author{Xiang S.W. Huang \textit{et al.}}
\email{mnakano@shibaura-it.ac.jp}
\noaffiliation

\maketitle

This Supplemental Material includes Sections~A--E, Figs.~S1--S11, and Table~S1.

\section{Superstructure-selective epitaxy of \texorpdfstring{\hcn{}}{Cr-NbSe2} thin films}

\begin{figure*}[t]
\includegraphics[width=\textwidth]{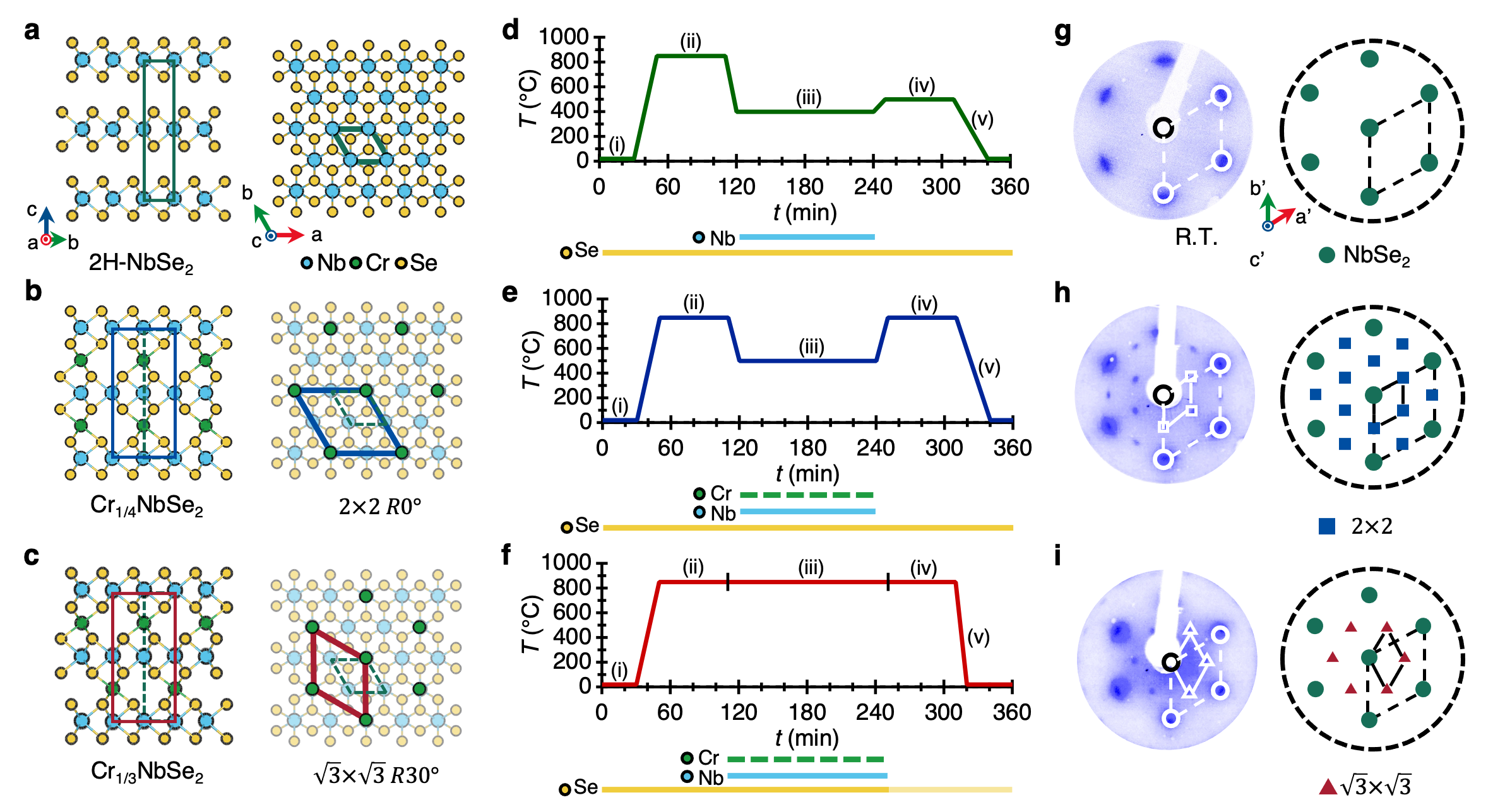}
\caption{\textbf{Superstructure-selective epitaxy of \cn{} thin films.}
\textbf{a-c,} Schematic crystal structures of (\textbf{a}) 2H-\nbse{}, (\textbf{b}) \cnq{}, and (\textbf{c}) \cnt{}. The blue, green, and yellow circles represent Nb, Cr, and Se atoms, respectively. The green rhombus in \textbf{a-c} represents the unit cell of \nbse{}, while the blue rhombus in \textbf{b} represents the \sqq{} superstructure in \cnq{} and the red rhombus in \textbf{c} represents the \sqt{} superstructure in \cnt{}, respectively. \textbf{d-f,} The growth sequence for (\textbf{d}) \nbse{}, (\textbf{e}) \cnq{}, and (\textbf{f}) \cnt{} epitaxial films. Each sequence consists of five steps: (i) Se deposition, (ii) Se treatment, (iii) main growth, (iv) post-growth annealing, (v) cooling. The color lines below the sequence represent the supply period of Cr (green: $\sim$0.05~\AA/s), Nb (sky blue: $\sim$0.05~\AA/s), and Se (yellow: $\sim$2~\AA/s; light yellow: $\sim$0.5~\AA/s). \textbf{g-i,} The LEED patterns and illustrations of the reciprocal lattice of (\textbf{g}) \nbse{}, (\textbf{h}) \cnq{}, and (\textbf{i}) \cnt{} epitaxial films. The $a^{*}$ and $b^{*}$ axes defined in \textbf{g} in reciprocal space are transformed from the $a$ and $b$ axes defined in \textbf{a} in real space. The dashed rhombus with round circle in \textbf{g-i} represents the unit cell of \nbse{}, while the solid rhombus with square in \textbf{h} represents the \sqq{} superstructure in \cnq{} and the solid rhombus with triangle in \textbf{i} represents the \sqt{} superstructure in \cnt{}, respectively.}
\label{figS1}
\end{figure*}

In this section, we summarize the MBE growth procedures used for \nbse{}, \cnq{}, and \cnt{} epitaxial thin films (Supplemental Fig.~\ref{figS1}). The growth follows established MBE and van der Waals epitaxy approaches for layered chalcogenides [1,2], but here the conditions are tuned to select the final Cr superstructure.

The growth sequence consists of five steps: (i) Se deposition, (ii) Se treatment, (iii) main growth, (iv) post-growth annealing, and (v) cooling. In step (i), the substrate surface is covered by amorphous Se at room temperature. The sample is then heated to 850~$^\circ$C and annealed under Se flux in step (ii), which cleans and reconstructs the substrate surface. After this treatment, the substrate temperature is adjusted to the growth temperature. Typical values are approximately 400~$^\circ$C for \nbse{}, 500~$^\circ$C for \cnq{}, and 850~$^\circ$C for \cnt{}.

The main growth in step (iii) is initiated by supplying Nb from the electron-beam evaporator. A typical growth rate is about 15~min per \nbse{} layer. For the Cr-intercalated phases, Cr is introduced during the growth of each \nbse{} layer by opening and closing the Knudsen-cell shutter. For \cnq{}, the Cr shutter is typically opened for 4~min within every 15~min growth cycle (i.e. opened for 4~min and closed for 11~min). For \cnt{}, the Cr shutter is opened for 5~min 20~s within every 15~min cycle.

After the main growth, the films are subjected to post-growth annealing in step (iv) to improve crystallinity and complete structural reconstruction. The annealing temperature is typically $\sim$500~$^\circ$C for \nbse{} and $\sim$850~$^\circ$C for \cnq{} and \cnt{}. During this step, the Se flux is maintained at $\sim$2~\AA/s for \nbse{} and \cnq{}, while it is reduced to $\sim$0.5~\AA/s for \cnt{}.

Finally, the samples are cooled to room temperature in step (v). \nbse{} and \cnq{} are typically cooled gradually over $\sim$30~min, whereas \cnt{} is obtained by rapid cooling within $\sim$5~min. RHEED shows that the superstructure develops during the cooling stage, indicating that the final Cr ordering is selected by the full thermal history rather than by the nominal Cr/Nb ratio alone. After growth, the films are capped with an amorphous Se layer of thickness $\sim$1~$\mu$m for protection during transfer in air.

\section{Determination of the in-plane lattice parameters of \texorpdfstring{\hcn{}}{Cr-NbSe2} thin films by RHEED}

\begin{figure*}[t]
\includegraphics[width=\textwidth]{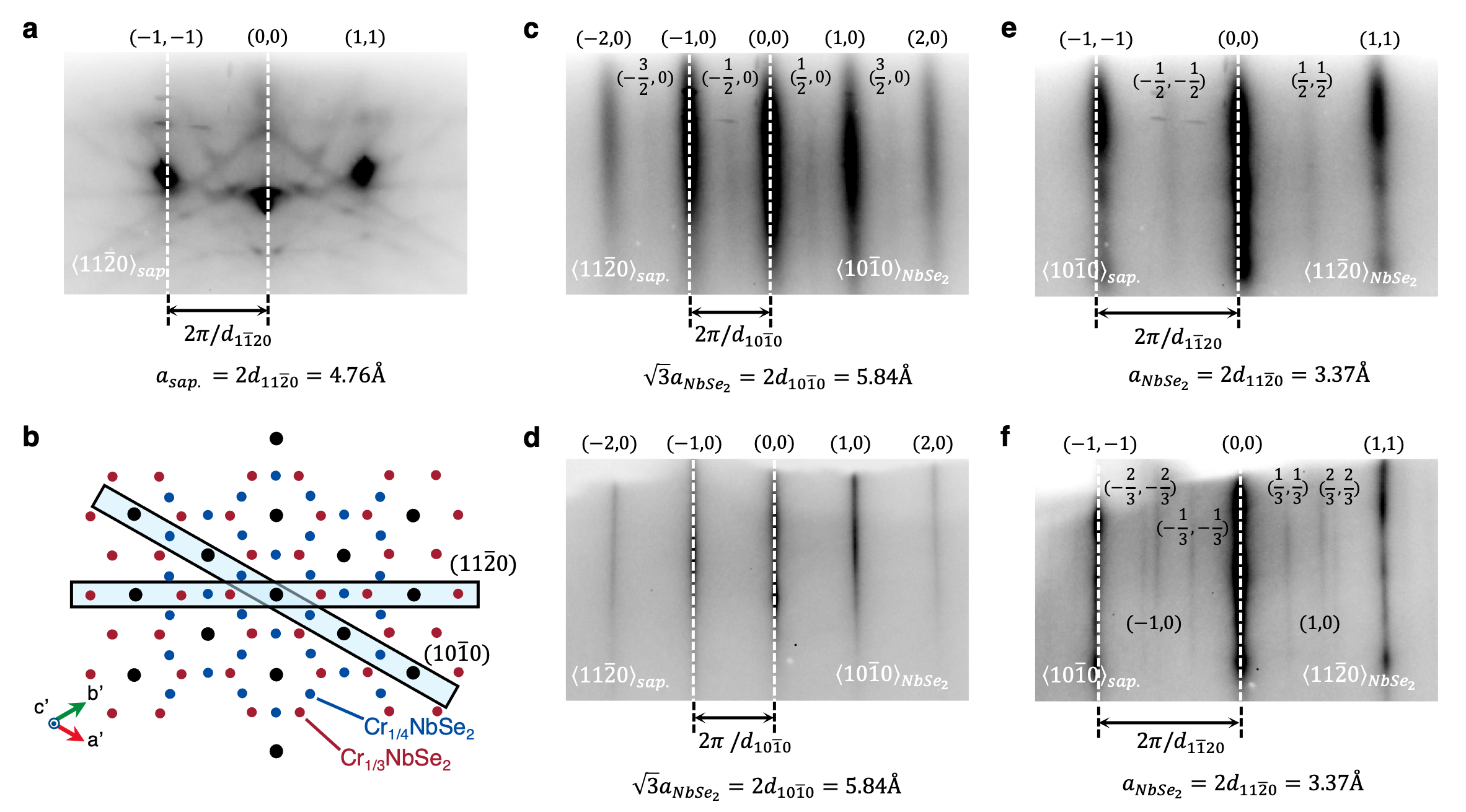}
\caption{\textbf{Determination of the in-plane lattice parameters of \cn{} thin films by RHEED.}
\textbf{a}, The RHEED pattern of the c-plane sapphire substrate measured along the $\langle 11\bar{2}0 \rangle$ azimuth for calibrating the reciprocal-space spacing. \textbf{b}, Schematic illustration of the reciprocal lattice of \nbse{} (black), \cnq{} (black and blue), and \cnt{} (black and red). The light blue rectangular area labelled with $(11\bar{2}0)$ and $(10\bar{1}0)$ includes the reciprocal lattice spots that satisfy the Bragg condition when the incident electron beam is irradiated along the $\langle 11\bar{2}0 \rangle$ and $\langle 10\bar{1}0 \rangle$ direction, respectively. \textbf{c-f}, The corresponding RHEED patterns of the (\textbf{c,e}) \cnq{} and (\textbf{d,f}) \cnt{} epitaxial films taken along (\textbf{c,d}) $\langle 11\bar{2}0 \rangle$ and (\textbf{e,f}) $\langle 10\bar{1}0 \rangle$ azimuth of the substrate. Note that the in-plane orientations of the obtained films were found to be rotated by 30$^\circ$ with respect to the substrate, and therefore, the azimuth angles are also rotated by 30$^\circ$ with respect to it. The \sqq{} superstructure pattern in \cnq{} was observed along both directions (\textbf{c,e}), whereas the \sqt{} pattern in \cnt{} was observed only along the $\langle 10\bar{1}0 \rangle$ direction of the substrate (and hence $\langle 11\bar{2}0 \rangle$ direction of the film) (\textbf{f}), which is consistent to the reciprocal lattice of each compound as shown in \textbf{b}. From the spacings of the RHEED streaks, the $a$-axis length of the host \nbse{} layer was calculated to be 3.37~\AA, providing the Cr-Cr distance of 6.74~\AA{} for \cnq{} and 5.84~\AA{} for \cnt{}, respectively.}
\label{figS2}
\end{figure*}

The in-plane lattice parameters of the obtained films were characterized by RHEED (Supplemental Fig.~\ref{figS2}). Supplemental Fig.~\ref{figS2}a shows the RHEED pattern of the c-plane sapphire substrate measured along the $\langle 11\bar{2}0 \rangle$ azimuth. The spacing between the principal diffraction features was used to calibrate the reciprocal-space scale, giving $a_{\mathrm{Sap}} = 4.76$~\AA.

Supplemental Fig.~\ref{figS2}b summarizes the reciprocal-lattice relations of \nbse{}, \cnq{}, and \cnt{}, together with the two reciprocal-space cuts probed in the RHEED experiment, providing the reciprocal-space counterpart to the real-space TEM measurements in the main text. The corresponding RHEED patterns are shown in Supplemental Figs.~\ref{figS2}c-f. The films are rotated by 30$^\circ$ with respect to the sapphire substrate, and therefore the azimuthal relation between substrate and film must be considered when indexing the streaks.

From the spacing of the original \nbse{} streaks, the host lattice constant is estimated as $a_{\mathrm{NbSe_2}} = 3.37$~\AA. This yields Cr-Cr distances of $2a = 6.74$~\AA{} for the \sqq{} \cnq{} phase and $\sqrt{3}\,a = 5.84$~\AA{} for the \sqt{} \cnt{} phase. The \sqq{} phase is identified by the additional half-order streaks visible in Supplemental Figs.~\ref{figS2}c and \ref{figS2}e, whereas the \sqt{} phase is identified by the one-third-order streaks visible in Supplemental Fig.~\ref{figS2}f. The absence of corresponding extra streaks in Supplemental Fig.~\ref{figS2}d along the other azimuth is consistent with the reciprocal-space selection rules of the \sqt{} ordering.

Out-of-plane structural characterization by x-ray diffraction is presented in the main text (Fig.~1i), where the 0002 reflections and clear Laue fringes confirm coherent out-of-plane growth and uniform thickness of the films.

\section{Magnetic properties of bulk \texorpdfstring{\hcnq{}}{Cr1/4NbSe2} single crystals}

\begin{figure*}[t]
\includegraphics[width=\textwidth]{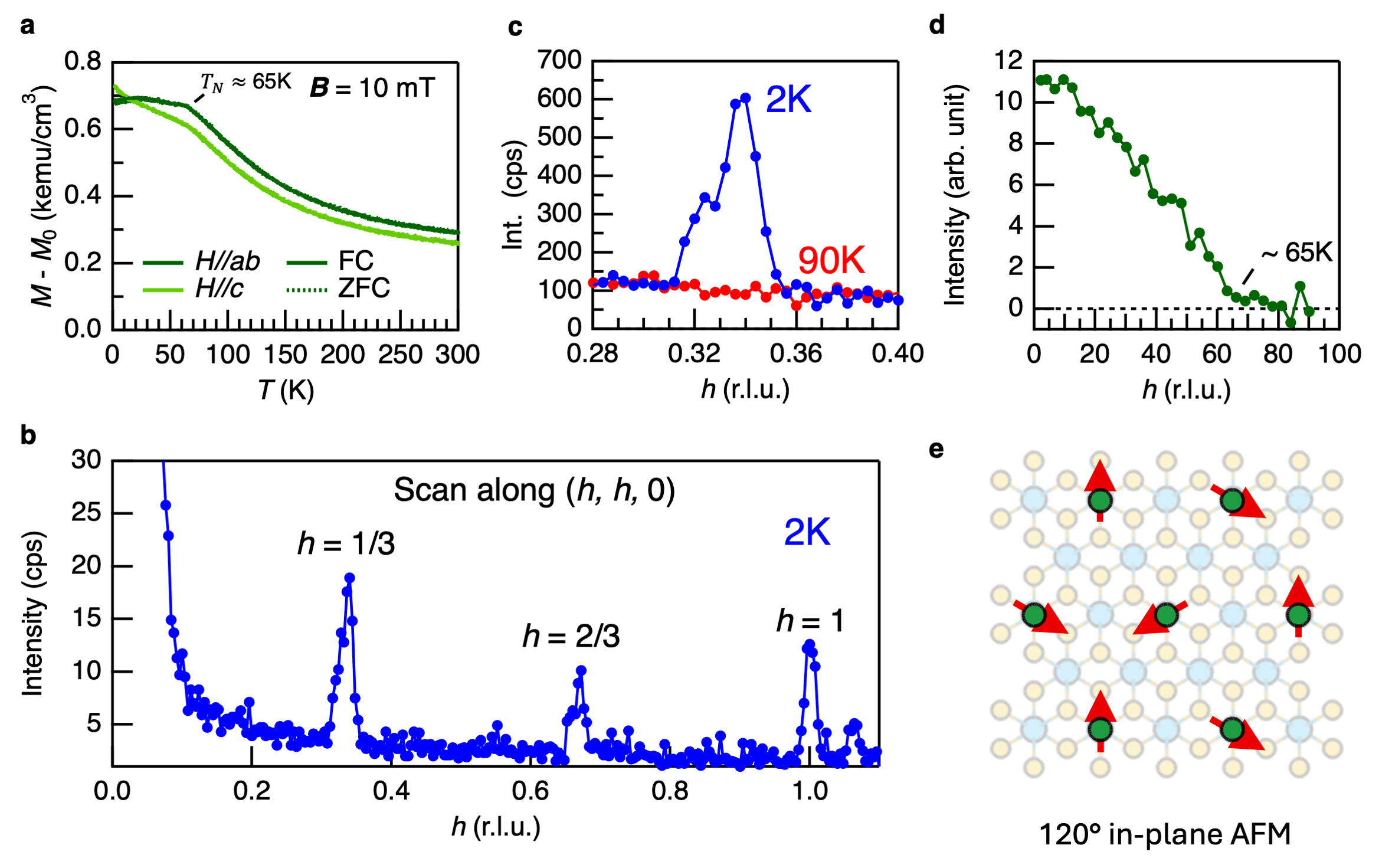}
\caption{\textbf{Magnetic properties of bulk \cnq{} single crystals.}
\textbf{a}, Temperature-dependent magnetization ($M - M_{0}$) at $B$ = 10~mT for $H$//ab and $H$//c under field-cooled (solid) and zero-field-cooled (dotted) conditions (almost overlapped), showing $T_{N} \approx 65$~K with no FC-ZFC splitting. \textbf{b}, Neutron scattering intensity along ($h$, $h$, 0) at 2~K. \textbf{c,} The $h$ = 1/3 magnetic reflection at 2~K and 90~K. \textbf{d,} Temperature dependence of the $h$ = 1/3 peak intensity. \textbf{e,} Schematic of the 120$^\circ$ in-plane antiferromagnetic structure identified by the polarized neutron diffraction experiments. Adapted from Yamaoka et al. [3].}
\label{figS3}
\end{figure*}

We synthesized bulk \cnq{} single crystals with the \sqq{} superstructure by chemical vapor transport (CVT) as a reference for the thin-film samples and companion study (Yamaoka et al. [3]). Figure~\ref{figS3}a shows the temperature-dependent magnetization after subtracting the temperature-independent component $M_{0}$. The bulk response closely resembles that of the 370L film, exhibiting a N\'eel-type transition at $T_{N} \approx 65$~K with the $H$//ab susceptibility substantially larger than $H$//c, confirming the easy-plane antiferromagnetic character.

To characterize the magnetic order periodicity of the bulk \cnq{}, a series of neutron scattering measurements were performed on the same CVT-grown crystals. The principal results are reported in Yamaoka et al. [3]. Neutron scattering along ($h$, $h$, 0) at 2~K (Fig.~\ref{figS3}b) reveals magnetic Bragg reflections at $h$ = 1/3, 2/3, confirming a $3\times3$ magnetic superstructure. The $h$ = 1/3 peak, present at 2~K, is absent at 90~K (Fig.~\ref{figS3}c), and its intensity vanishes at approximately 65~K (Fig.~\ref{figS3}d), consistent with $T_{N}$ identified by the magnetization measurement (Fig.~\ref{figS3}a). Polarized neutron scattering experiments identified the 120$^\circ$ in-plane antiferromagnetic ground state (Fig.~\ref{figS3}e), with the Heisenberg-like magnetic anisotropy.

We note that $T_{N} \approx 65$~K of the bulk sample is higher than that of the 370L thin-film sample ($\sim$50~K, main text Fig.~3a); such layer-dependent suppression of the ordering temperature is expected in finite-thickness films.

\section{Multi-carrier transport and anisotropic magnetoresistance in the 6L-thick \texorpdfstring{\hcnq{}}{Cr1/4NbSe2}}
\label{secD}

This section presents transport evidence supporting the low-carrier-density, multicarrier identification of the \cnq{} itinerant background discussed in the main text. We analyze the antisymmetrized Hall response and the magnetoresistance in three field-current geometries.

\subsection{Multi-carrier transport in the 6L-thick \texorpdfstring{\hcnq{}}{Cr1/4NbSe2}}

In a two-carrier system with partial conductivities $\sigma_{1}$, $\sigma_{2}$ and Hall mobilities $\mu_{1}$, $\mu_{2}$, the conductivity tensor components are
\begin{equation}
\sigma_{xx} = \frac{\sigma_{1}}{1 + \mu_{1}^{2}H^{2}} + \frac{\sigma_{2}}{1 + \mu_{2}^{2}H^{2}},
\label{eqS1}
\end{equation}
\begin{equation}
\sigma_{xy} = \frac{\sigma_{1}\mu_{1}H}{1 + \mu_{1}^{2}H^{2}} + \frac{\sigma_{2}\mu_{2}H}{1 + \mu_{2}^{2}H^{2}}.
\label{eqS2}
\end{equation}
The Hall resistivity follows as $\rho_{xy} = \sigma_{xy}/(\sigma_{xx}^{2} + \sigma_{xy}^{2})$. Expanding in powers of $H$ at low fields ($\mu_{i}H \ll 1$) and retaining terms to third order gives
\begin{equation}
\rho_{xy} = AH + CH^{3},
\label{eqS3}
\end{equation}
where
\begin{equation}
A = \frac{\sigma_{1}\mu_{1} + \sigma_{2}\mu_{2}}{(\sigma_{1} + \sigma_{2})^{2}},
\label{eqS4}
\end{equation}
\begin{equation}
C = - \frac{\sigma_{1}\sigma_{2}(\mu_{1} - \mu_{2})^{2}(\sigma_{1}\mu_{2} + \sigma_{2}\mu_{1})}{(\sigma_{1} + \sigma_{2})^{4}}.
\label{eqS5}
\end{equation}
$C$ vanishes identically when $\mu_{1} = \mu_{2}$, recovering the single-carrier linear Hall effect [4]. A non-zero $C$ therefore constitutes a direct signature of multicarrier transport, independent of whether the two carriers have the same or opposite sign.

\begin{figure*}[t]
\includegraphics[width=\textwidth]{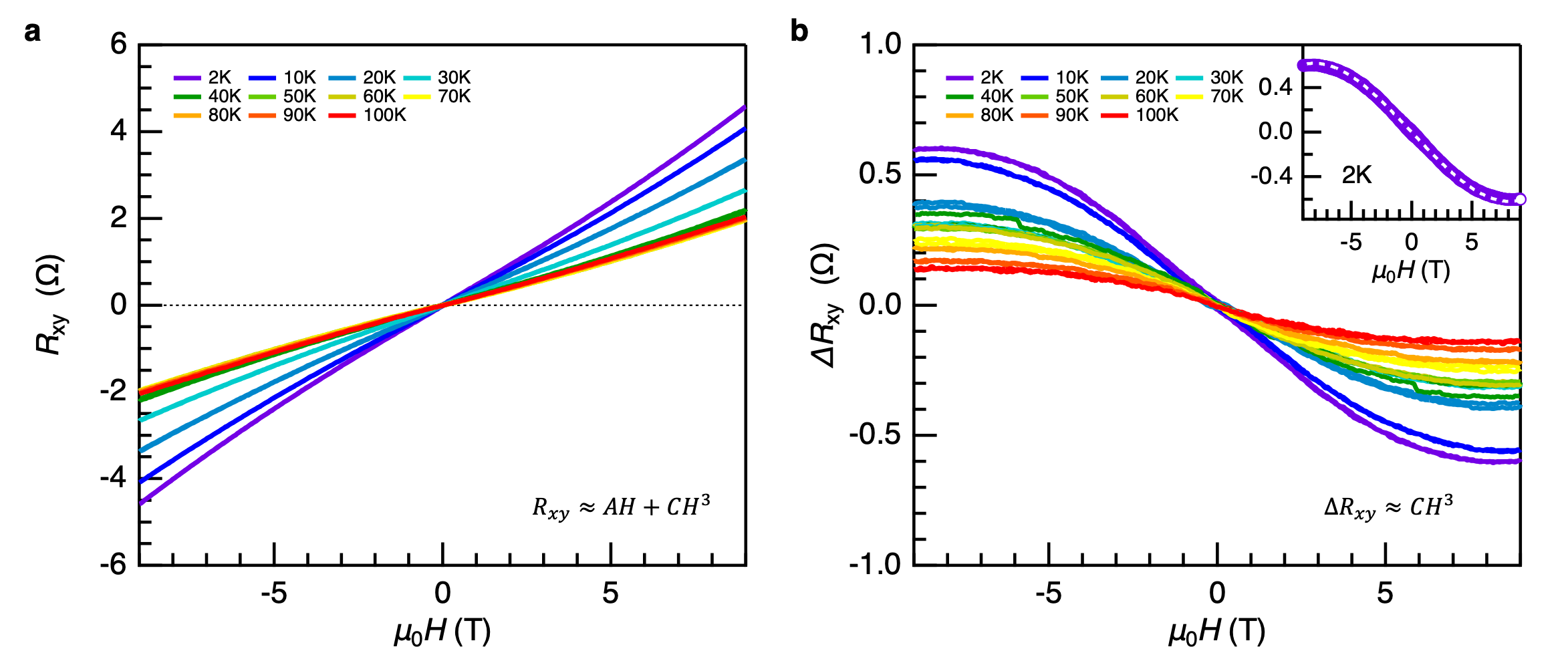}
\caption{\textbf{Multi-carrier transport in the 6L-thick \cnq{}.}
\textbf{a,} Antisymmetrized Hall response $R_{xy}(H)$ for $H$//$c$ at temperatures from 2~K to 100~K. \textbf{b,} Cubic component $\Delta R_{xy} \approx CH^{3}$ extracted at the same temperatures, plotted on an expanded vertical scale. Inset: representative fit of the 2~K curve to $\rho_{xy} \approx AH + CH^{3}$ (sign convention reflects the electron-leaning balance at this temperature). The non-zero cubic response across all measured temperatures directly establishes multi-band transport in the 6L film: a single-band Drude expression has $C \equiv 0$ identically.}
\label{figS4}
\end{figure*}

We apply this analysis to the as-grown 6L \cnq{} film. Figure~\ref{figS4}a shows the antisymmetrized Hall resistance $R_{xy}(H)$ at representative temperatures. Figure~\ref{figS4}b shows the residual after subtracting the linear component $AH$, isolating the cubic contribution $CH^{3}$. The cubic term is clearly resolved across all measured temperatures, confirming multicarrier transport. Table~\ref{tabS1} lists the fitted $A$ and $C$ coefficients at different temperatures.

The close agreement between the cubic fit and the measured $R_{xy}(H)$ validates the multicarrier origin of the observed non-linear Hall response.

\subsection{Anisotropic magnetoresistance in the 6L-thick \texorpdfstring{\hcnq{}}{Cr1/4NbSe2}}

\begin{figure*}[t]
\includegraphics[width=\textwidth]{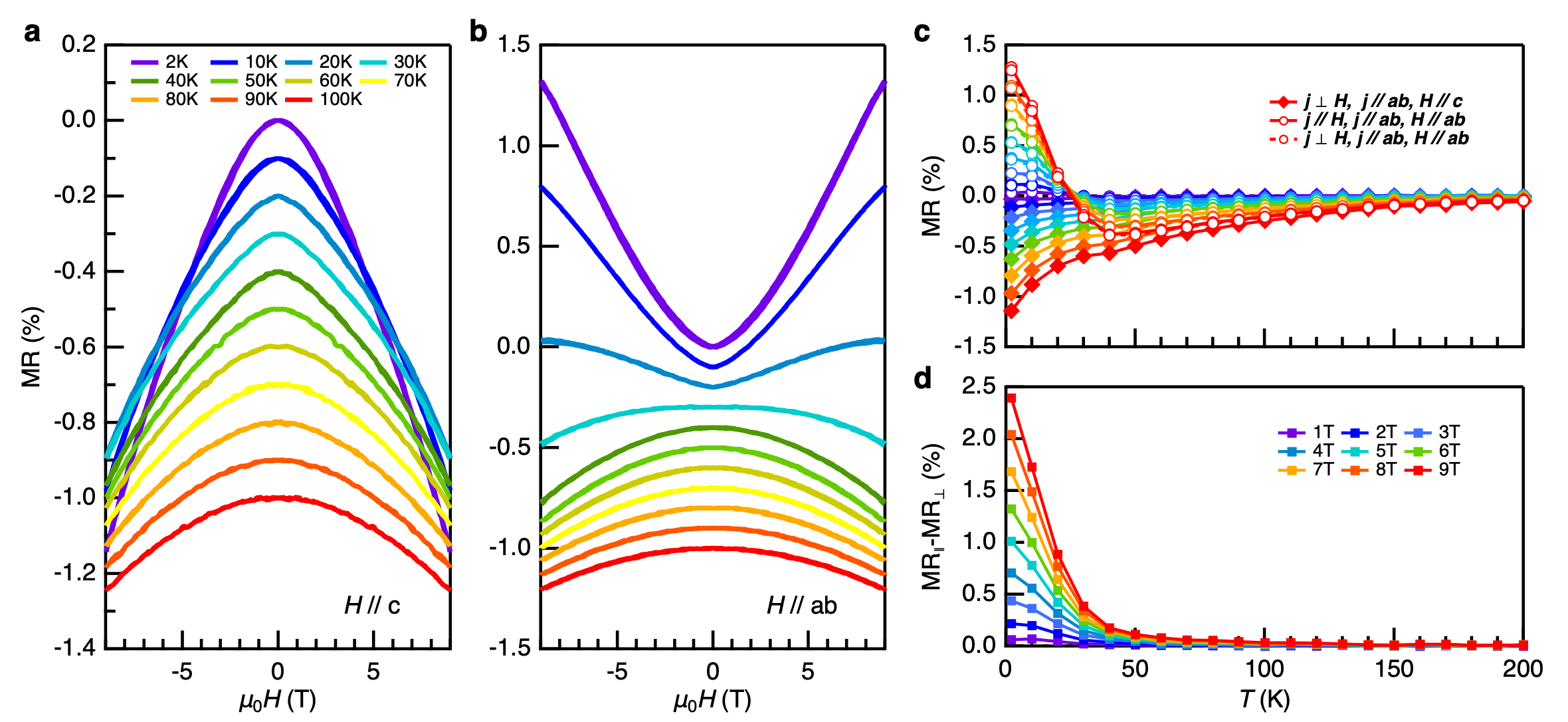}
\caption{\textbf{Anisotropic magnetoresistance in the 6L-thick \cnq{}.}
\textbf{a,} MR($H$) for $H$//$c$ at temperatures from 2~K to 100~K: negative across the full range. \textbf{b,} MR($H$) for $H$//$ab$: positive at low $T$ (2~K - 20~K) and negative above $\sim$30~K. \textbf{c,} Temperature dependence of MR for three field-current geometries: $j \perp H$ with $H$//$c$ (filled diamonds); $j$//$H$ with $H$//$ab$ (open circles, solid line); and $j \perp H$ with $H$//$ab$ (open circles, dashed line). The two in-plane configurations are identical, confirming in-plane isotropy. \textbf{d,} In-plane MR anisotropy MR$_{\parallel}$ $-$ MR$_{\perp}$ for fields 1~T - 9~T, vanishing above $T^{*} \approx 50$~K.}
\label{figS5}
\end{figure*}

Supplemental Figure~\ref{figS5} shows the magnetoresistance of the 6L \cnq{} film in three field-current geometries. The two in-plane configurations ($j$//$H$ and $j \perp H$, both with $H$//ab) yield identical responses, confirming in-plane isotropy consistent with an easy-plane spin configuration.

On the other hand, the $H$//ab and $H$//c responses differ qualitatively. MR($H$//ab) is positive at low temperature and changes sign to negative above $\sim$30~K, whereas MR($H$//c) remains negative across the full temperature range. This sign reversal excludes a purely orbital origin, identifying the MR as magnetic-scattering dominated.

The MR anisotropy between $H$//c and $H$//ab vanishes above $T^{*} \approx 50$~K, identifying $T^{*}$ as the upper boundary of magnetic correlations in the 6L film. This is the same scale at which $\rho(T)$ and $R_{H}(T)$ show kinks (main text Figs.~2i,k). In the 6L film, the absolute magnetic moment is too small to be detected in $M(T)$; the MR therefore provides the sole transport-based identification of the magnetic ordering scale for those ultrathin samples.

\section{Design of the post-growth annealing experiments}

\begin{figure*}[t]
\includegraphics[width=\textwidth]{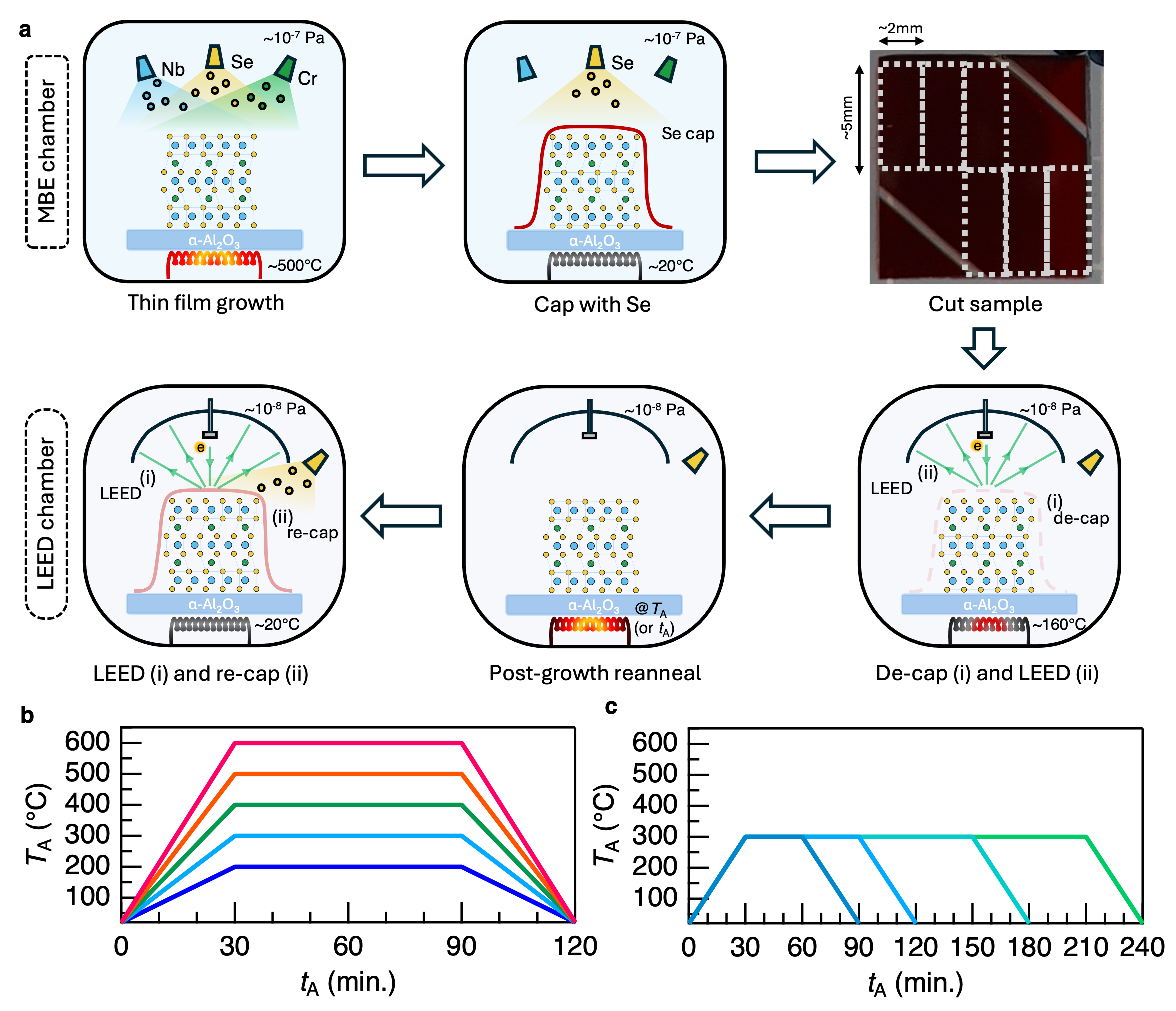}
\caption{\textbf{Design of the post-growth annealing experiments.}
\textbf{a}, Illustrations of the post-growth annealing process. \cnq{} film was grown on a 10~mm $\times$ 10~mm substrate and capped with Se for protection. The sample was then cut into 5~mm $\times$ 2~mm small pieces. Each piece was transferred into the vacuum chamber equipped with LEED, and (i) heated at $\sim$160~$^\circ$C for 1 hour to remove a Se capping layer (de-cap). After cooling down, (ii) the superstructure pattern was checked by LEED. Then, the sample was treated at $T_{A}$ for 1 hour. After this thermal treatment, (i) the superstructure pattern was checked again by LEED, followed by (ii) deposition of a Se capping layer (re-cap). The sample was then taken out of the LEED chamber and subjected to magnetization and transport measurements. \textbf{b},\textbf{c}, Typical annealing sequences for (\textbf{b}) $T_{A}$- and (\textbf{c}) $t_{A}$-dependence experiments.}
\label{figS6}
\end{figure*}

The distinct growth windows of \cnq{} ($\sim$500~$^\circ$C) and \cnt{} ($\sim$850~$^\circ$C) motivated us to ask whether the Cr superstructure of a grown film can be modified by subsequent thermal treatment. To test this systematically, we designed the following post-growth annealing experiments as schematically illustrated in Fig.~\ref{figS6}.

\begin{figure*}[t]
\includegraphics[width=\textwidth]{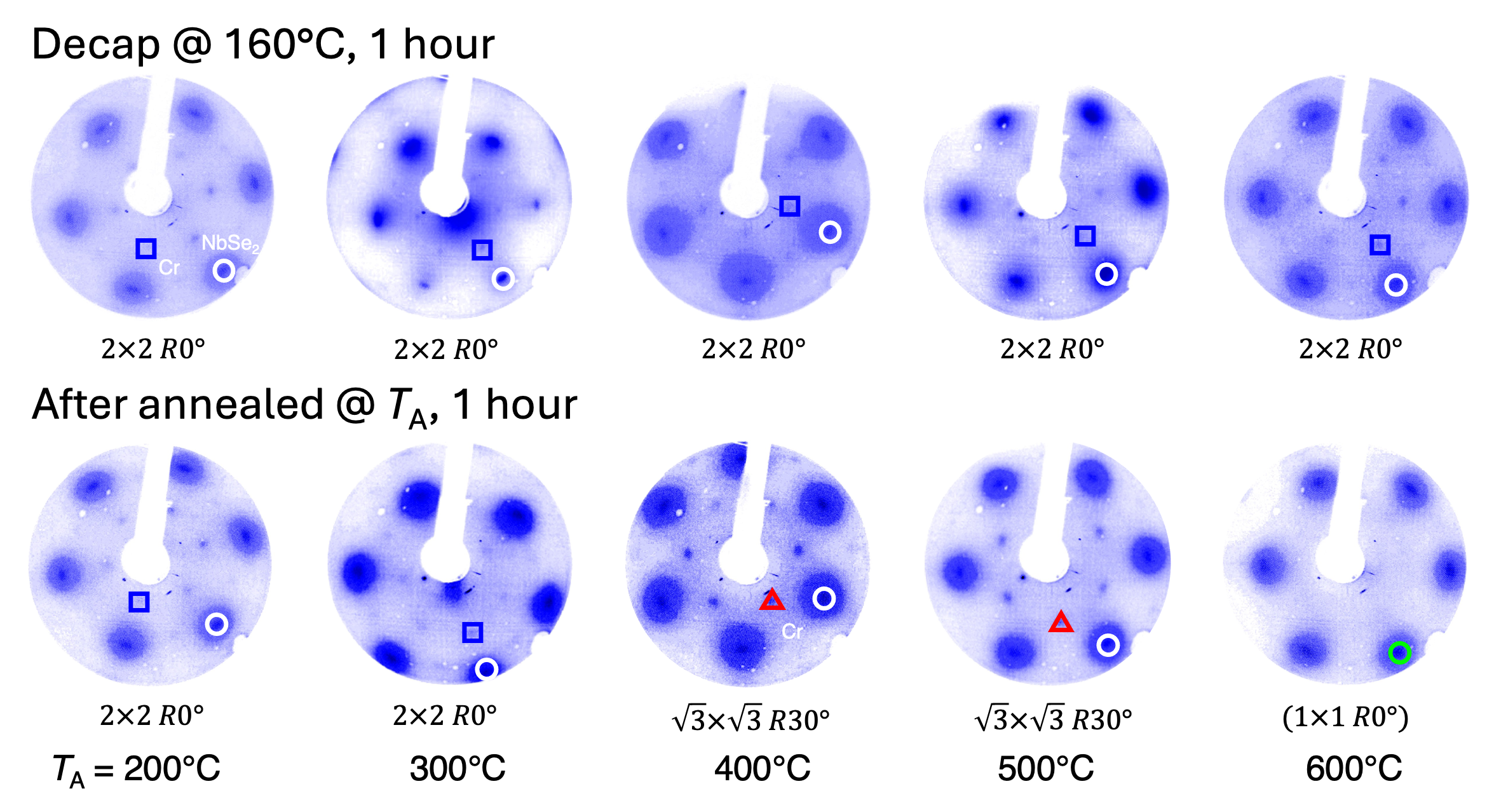}
\caption{\textbf{Evolutions of the LEED patterns in \cnq{} films upon the post-growth annealing at $T_{A}$ for $t_{A}$ = 1 hour.}
The upper panels show the LEED patterns of all the samples subjected to the annealing-temperature ($T_{A}$)-dependence experiments after the Se de-capping process (just before the post-growth annealing process), exhibiting the \sqq{} superstructures. The lower panels show the LEED patterns of the corresponding samples after the post-growth annealing process at specific temperature ($T_{A}$) for 1 hour (the data same as those shown in Fig.~4a in the main text).}
\label{figS7}
\end{figure*}

\begin{figure*}[t]
\includegraphics[width=\textwidth]{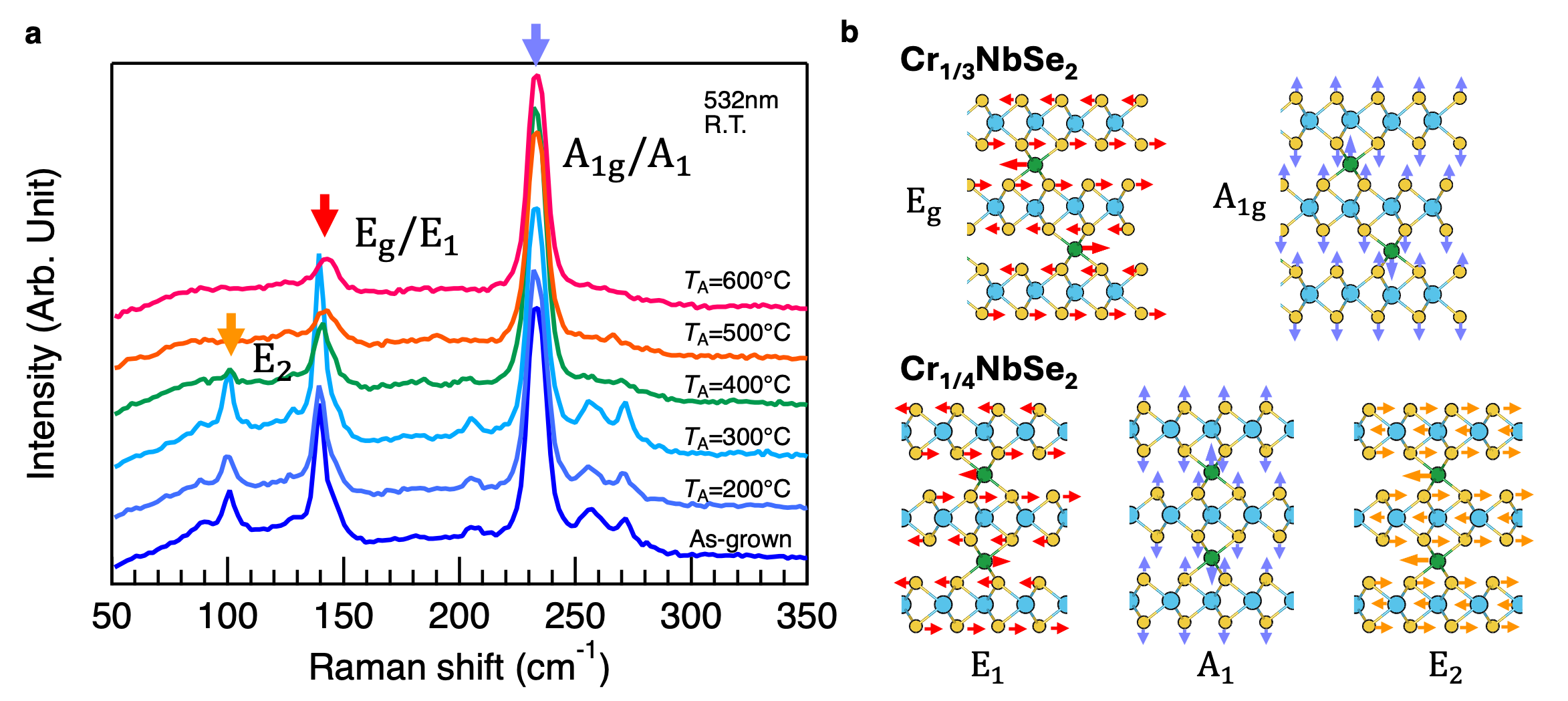}
\caption{\textbf{Evolutions of the Raman spectra of \cnq{} films upon the post-growth annealing at $T_{A}$ for $t_{A}$ = 1 hour.}
\textbf{a,} Room-temperature Raman spectra of \cnq{} films after annealing at different $T_{A}$, measured with a 532~nm excitation source. The spectra evolve systematically across the annealing series, with clear changes in the phonon features near the modes labelled $E_{2}$, $E_{g}/E_{1}$, and $A_{1g}/A_{1}$. \textbf{b,} Schematic displacement patterns of the representative Raman-active modes used for assignment in \cnt{} and \cnq{} [5]. Unlike LEED, which probes only the topmost layer of the film, Raman is sensitive to the full film thickness; the systematic evolution of the Raman spectra with $T_{A}$ therefore demonstrates that the annealing-induced structural change extends through the thickness of the film rather than being limited to the surface.}
\label{figS8}
\end{figure*}

We first grow \cnq{} film on a relatively large ($\sim$10~mm $\times$ 10~mm) $c$-plane sapphire substrate in the MBE chamber, followed by the formation of a Se capping layer for protection. The sample is then taken out of the chamber and cut into 5~mm $\times$ 2~mm small pieces. Each piece is then transferred into the vacuum chamber and subjected to the post-growth annealing process at specific temperature defined as $T_{A}$ under the ultrahigh-vacuum condition ($< 10^{-7}$~Pa). The sample is first heated at $\sim$160~$^\circ$C for 1 hour to remove Se capping layer (de-cap). After cooling down to room temperature, LEED confirms that the original \sqq{} superstructure is retained (Fig.~\ref{figS6}a). Then, the sample is treated at $T_{A}$ for 1 hour as the sequence in Fig.~\ref{figS6}b shows. After this thermal treatment, the superstructure is again checked by LEED, which evolves with $T_{A}$ as shown in Fig.~4a of the main text and Fig.~\ref{figS7}. After this LEED observation, a Se capping layer is again formed on to the sample surface (re-cap). The sample is then taken out of the chamber and subjected to magnetization and transport measurements. Raman spectroscopy was also performed on the same annealing series (Fig.~\ref{figS8}) and evolves consistently with the LEED-resolved structural change, confirming that the superstructure reorganization of \sqq{}-to-\sqt{} extends through the full film thickness instead of being limited to the surface [5].

\begin{figure*}[t]
\includegraphics[width=\textwidth]{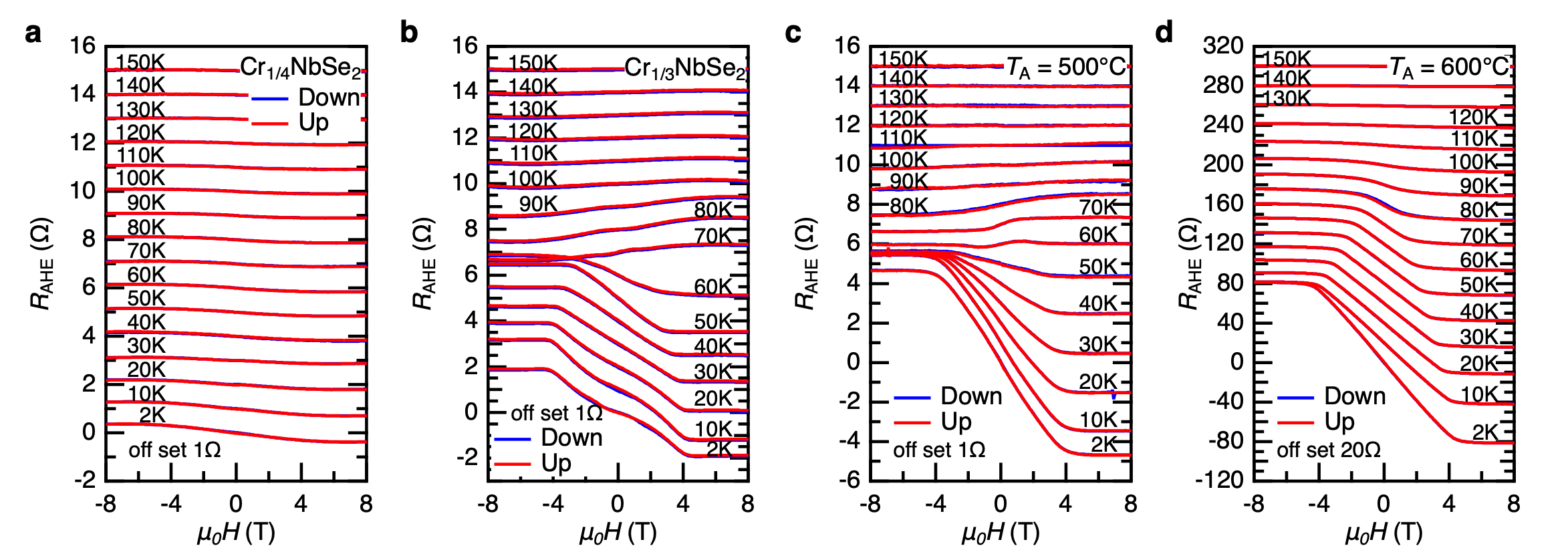}
\caption{\textbf{The anomalous Hall effect data of \cn{} epitaxial films.}
\textbf{a-d}, The anti-symmetrized anomalous Hall effect data of the (\textbf{a}) as-grown \cnq{}, (\textbf{b}) as-grown \cnt{}, (\textbf{c}) vacuum-annealed \cnq{} ($T_{A}$ = 500~$^\circ$C), and (\textbf{d}) vacuum-annealed \cnq{} ($T_{A}$ = 600~$^\circ$C) epitaxial films measured at different temperatures. The ordinary Hall components proportional to the external magnetic field were already subtracted.}
\label{figS9}
\end{figure*}

The 500~$^\circ$C and 600~$^\circ$C samples exhibit a saturating anomalous Hall component tracking $M(H)$, confirming ferromagnetic order; no such component is detected for the as-grown through 400~$^\circ$C samples (Fig.~\ref{figS9}). The non-linear Hall signals of the non-ferromagnetic samples are analyzed within the framework of the multi-band model in Supplemental Material Section~\ref{secD}.

\begin{figure*}[t]
\includegraphics[width=\textwidth]{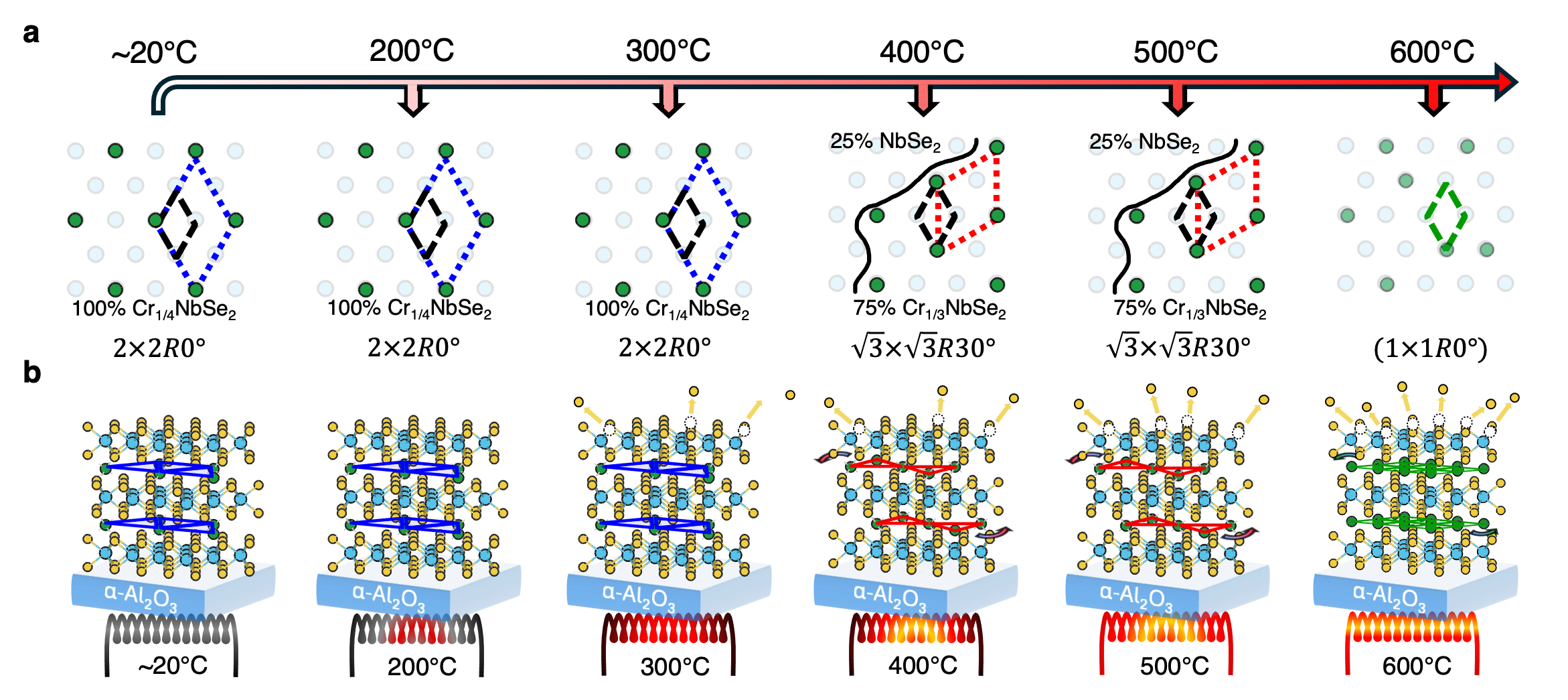}
\caption{\textbf{Proposed effects of the post-growth annealing process on the structural properties of \cnq{} epitaxial films.}
\textbf{a}, The evolution of the in-plane Cr superstructure in \cnq{} upon the post-growth annealing process under the ultra-high-vacuum condition ($< 10^{-7}$~Pa). Assuming that the Cr amount does not change during this process, the transformation of the superstructure pattern from \sqq{} to \sqt{} should accompany a phase separation from 100\% of \cnq{} to a mixture of 75\% of \cnt{} and 25\% of \nbse{}, which is consistent with the magnetization data shown in Fig.~4b in the main text. \textbf{b}, Illustrations of the changes in \cnq{} films upon the post-growth annealing process. Se starts to evaporate from the host \nbse{} layers above $T_{A}$ = 300~$^\circ$C, leading to the increase of the amount of Se deficiency. The superstructure transformed from \sqq{} (blue) to \sqt{} (red), and finally vanished (green). The formation of Se vacancy is considered to induce free electrons in the system, which is consistent with the Hall-effect data shown in Fig.~4d in the main text.}
\label{figS10}
\end{figure*}

We summarize possible effects of the post-growth annealing process on the structural properties of \cnq{} epitaxial film in Fig.~\ref{figS10}.


\begin{table}[t]
\caption{\textbf{Multi-band fit for the Hall response in the as-grown 6L-thick \cnq{}.} Fit form: $\rho_{xy}(H) = AH + CH^{3}$ applied to antisymmetrized Hall data from $-9$~T to 9~T at each temperature. Sheet units ($\Omega$/T and $\Omega$/T$^{3}$ respectively). $R^{2}$ is the coefficient of determination.}
\label{tabS1}
\begin{ruledtabular}
\begin{tabular}{cccc}
$T$ (K) & $A$ ($\Omega$/T) & $C$ ($\Omega$/T$^{3}$) & $R^{2}$ \\
\hline
2   & 0.4625 & $5.980\times10^{-4}$ & 0.99999 \\
10  & 0.4146 & $4.962\times10^{-4}$ & 0.99998 \\
20  & 0.3455 & $3.617\times10^{-4}$ & 0.99996 \\
30  & 0.2719 & $2.992\times10^{-4}$ & 0.99997 \\
40  & 0.2164 & $3.323\times10^{-4}$ & 0.99994 \\
50  & 0.1983 & $3.541\times10^{-4}$ & 0.99989 \\
60  & 0.1935 & $3.425\times10^{-4}$ & 0.99984 \\
70  & 0.1953 & $3.172\times10^{-4}$ & 0.99975 \\
80  & 0.2002 & $2.771\times10^{-4}$ & 0.99983 \\
90  & 0.2056 & $2.388\times10^{-4}$ & 0.99985 \\
100 & 0.2108 & $2.134\times10^{-4}$ & 0.99989 \\
110 & 0.2163 & $1.869\times10^{-4}$ & 0.99991 \\
120 & 0.2211 & $1.632\times10^{-4}$ & 0.99995 \\
130 & 0.2258 & $1.458\times10^{-4}$ & 0.99993 \\
140 & 0.2319 & $1.133\times10^{-4}$ & 0.99994 \\
150 & 0.2369 & $1.025\times10^{-4}$ & 0.99994 \\
160 & 0.2442 & $0.607\times10^{-4}$ & 0.99997 \\
170 & 0.2443 & $0.700\times10^{-4}$ & 0.99994 \\
180 & 0.2463 & $0.740\times10^{-4}$ & 0.99996 \\
190 & 0.2482 & $0.804\times10^{-4}$ & 0.99993 \\
200 & 0.2515 & $0.620\times10^{-4}$ & 0.99991 \\
\end{tabular}
\end{ruledtabular}
\end{table}


\begin{figure}[t]
\includegraphics[width=\columnwidth]{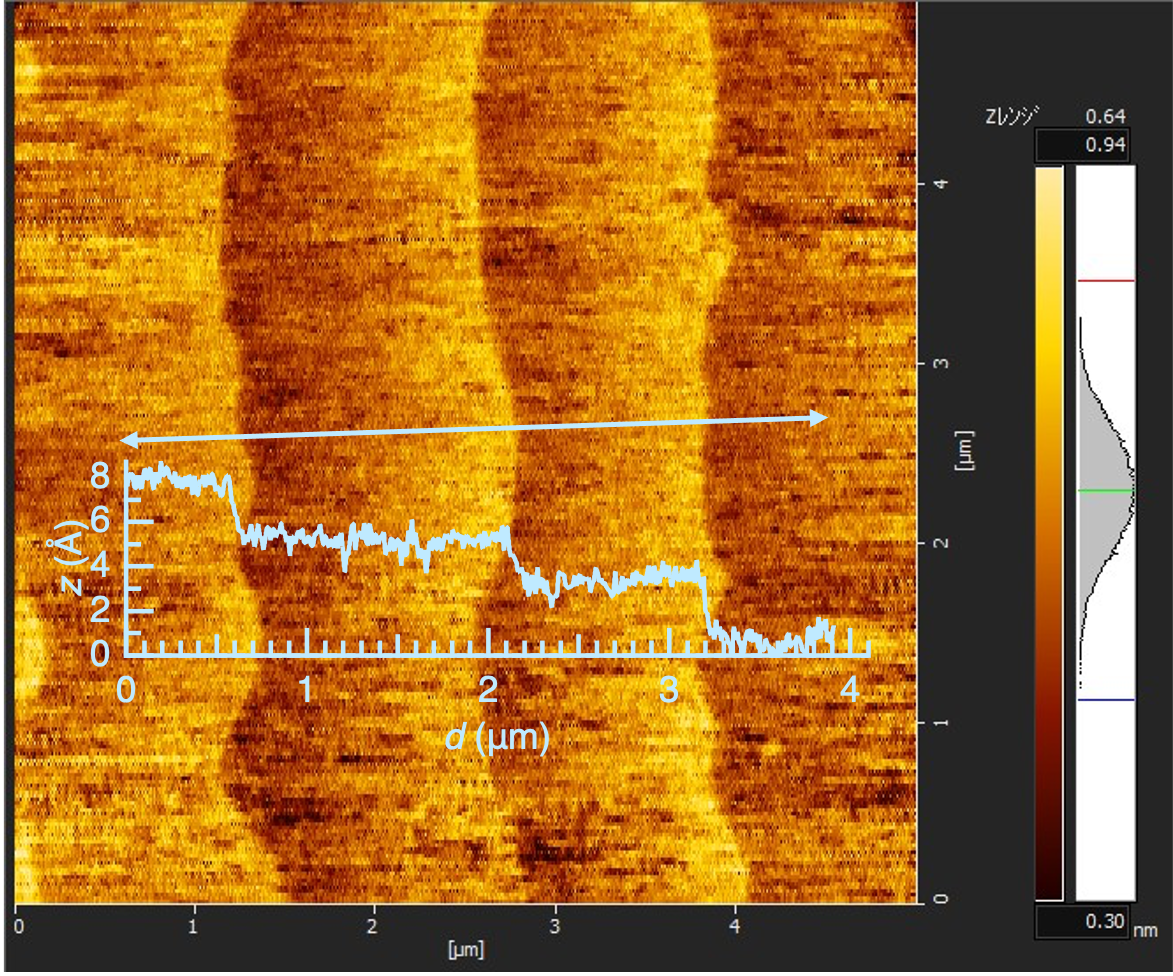}
\caption{\textbf{Surface morphology of the air-annealed $c$-plane sapphire substrate.}
A typical surface morphology of a $c$-plane sapphire substrate annealed at 1000~$^\circ$C for 3 hours in air characterized by atomic-force microscopy, showing the atomically-flat surface with the step and terrace structure. The inset shows the cross-sectional profile along the off-angle direction, showing a terrace width $\sim$1~$\mu$m and a step height $\sim$3~\AA.}
\label{figS11}
\end{figure}
